\documentclass[11pt,a4paper]{article}

\usepackage{jheppub}
\usepackage{graphicx}
\usepackage{dcolumn}
\usepackage{bm}
\usepackage{amsmath}
\usepackage{braket}
\usepackage{slashed}
\usepackage{epstopdf}
\usepackage{placeins}
\usepackage{multirow}
\usepackage{makecell}
\usepackage{soul}
\usepackage{braket}
\usepackage[shortlabels]{enumitem}
\usepackage{placeins}
\usepackage[font=small, labelfont=bf, textfont={small,it}]{caption}
\newcommand{\beq}{\begin{eqnarray}}
\newcommand{\eeq}{\end{eqnarray}}
\newcommand{\beqnn}{\begin{eqnarray*}}
\newcommand{\eeqnn}{\end{eqnarray*}}

\renewcommand{\thesection}{\arabic{section}}

\newcommand{\Tr}{\ensuremath{\mathrm{Tr}}}

\newcommand{\YM}{\ensuremath{\mathrm{YM}}}
\newcommand{\stag}{\ensuremath{\mathrm{stag}}}
\newcommand{\SP}{\ensuremath{\mathrm{SP}}}

\newcommand{\stout}{\ensuremath{\mathrm{stout}}}

\newcommand{\SU}{\ensuremath{\mathrm{SU}}}

\newcommand{\LQCD}{\ensuremath{\mathrm{LQCD}}}

\newcommand{\R}{\ensuremath{\mathrm{R}}}
\newcommand{\phys}{(\ensuremath{\mathrm{phys}})}

\begin{document}
	
\title{The chiral condensate of $N_f=2+1$ QCD from the spectrum of the staggered Dirac operator}

\author[a]{Claudio Bonanno,}
\author[b]{Francesco D'Angelo}
\author[b]{and Massimo D'Elia}

\affiliation[a]{Instituto de F\'isica Te\'orica UAM-CSIC, c/ Nicol\'as Cabrera 13-15, Universidad Aut\'onoma de Madrid, Cantoblanco, E-28049 Madrid, Spain}

\affiliation[b]{Università di Pisa and INFN Sezione di Pisa, Largo B.~Pontecorvo 3, I-56127 Pisa, Italy}

\emailAdd{claudio.bonanno@csic.es}
\emailAdd{francesco.dangelo@phd.unipi.it}
\emailAdd{massimo.delia@unipi.it}

\abstract{
\noindent We compute the chiral condensate of $2+1$ QCD from the mode number of the staggered Dirac operator, performing controlled extrapolations to both the continuum and the chiral limit. We consider also alternative strategies, based on the quark mass dependence of the topological susceptibility and of the pion mass, and obtain consistent results within errors. Results are also consistent with phenomenological expectations and with previous 
numerical determinations obtained with different lattice discretizations.}


\maketitle

\flushbottom

\section{Introduction}\label{sec:intro}

Flavor symmetries in QCD and their realization play a key role in determining the most relevant theoretical non-perturbative features of the theory, as well as being linked to several interesting phenomenological aspects of strong interactions. In this respect, Chiral Perturbation Theory~\cite{Gasser:1983yg,Gasser:1984gg} (ChPT) is able to provide remarkably accurate qualitative and quantitative predictions about QCD from an effective Lagrangian essentially built over the salient properties of the chiral symmetry exhibited by the full theory, once a few Low Energy Constants (LECs), not computable within the effective theory alone, are fixed. A possible way to fix them is to match ChPT predictions with experimentally measurable quantities (such as hadron masses). Another, more theoretically driven, strategy is to match such quantities with observables of the full theory, and to compute them non-perturbatively from first-principles methods, such as lattice Monte Carlo simulations.

At Leading Order (LO) of the chiral expansion, and considering only two light quark flavors $(u,d)$, only two LECs are needed to fully specify the $\SU(2)$ effective theory, the chiral condensate $\Sigma$ and the pion decay constant $F_\pi$:
\beq\label{eq:Sigma_def}
\Sigma \equiv - \lim_{m_u,m_d\to 0} \braket{\overline{u} u},
\eeq
\beq\label{eq:Fpi_def}
F_\pi \equiv \lim_{m_u,m_d\to 0} \frac{1}{M_\pi}\bra{\Omega} \overline{u}\gamma_4\gamma_5 d \ket{\pi(\vec{p}=\vec{0})},
\eeq
where $M_\pi$ is the pion mass, and $u$ and $d$ represent, respectively, the spinors of the up and down quarks.

Phenomenological estimates can be obtained combining experimental results and theoretical ChPT predictions, in particular the well-known Gell-Mann--Oakes--Renner (GMOR) relation, which, at LO in ChPT with 2 light flavors, reads:
\beq\label{eq:GMOR_nondegenerate}
M^2_\pi = \frac{\Sigma}{F^2_\pi} (m_u + m_d).
\eeq
The PDG~\cite{Workman:2022ynf} reports: $M_\pi = 134.9768(5)$~MeV, $F_\pi^{\phys} = 92.28(11)$~MeV\footnote{In this work we follow the convention which yields $F_\pi = F_\pi^{\phys} \simeq 92$~MeV at the physical point. Other authors adopt the equivalent convention $f_\pi = \sqrt{2} F_\pi$, leading to $f_\pi^{\phys} \simeq 130$~MeV.}, $m_u^{\phys} = 2.16(49)$~MeV and $m_d^{\phys} = 4.67(48)$~MeV, leading to $\Sigma_{\mathrm{pheno}} = [283(24)~\text{MeV}]^3$. Although such estimation is done using the pion decay constant at the physical point, while the GMOR relation involves the pion decay constant in the chiral limit, cf.~Eq.~\eqref{eq:Fpi_def}, this number turns out to be in perfect agreement with lattice calculations, as we will discuss in the following. This can be viewed as a confirmation of the main hypothesis underlying ChPT, i.e., that the $u$ and $d$ quark masses are sufficiently light to be treated as a small perturbation of the massless theory.

Generally speaking, chiral symmetry is a delicate issue from the viewpoint of lattice simulations. As a matter of fact, on one hand, chiral-symmetry preserving fermionic discretizations, such as Domain Wall or Overlap, are extremely computationally demanding, especially for a calculation targeting a few percent accuracy. On the other hand, the most typically employed quark discretizations, namely Wilson and staggered quarks, are computationally cheaper but explicitly break, either completely or partially, the chiral symmetry, which  is recovered only in the continuum limit. This issue can introduce non-trivial numerical challenges, as it is well known, e.g., from lattice calculations of the topological susceptibility, where explicit chiral symmetry breaking and the absence of exact zero modes of the Dirac operator are the dominant sources of lattice artifacts~\cite{Bonati:2015vqz,Petreczky:2016vrs,Borsanyi:2016ksw,Frison:2016vuc,Alexandrou:2017bzk,Bonati:2018blm,Athenodorou:2022aay}.

Despite such issues, however, in the last ten years many lattice determinations adopting both chiral and non-chiral quarks of these and other QCD LECs have appeared in the literature, both for $2$~\cite{ETM:2009ztk,Cichy:2013gja,Brandt:2013dua,Engel:2014eea}, $2+1$~\cite{Bazavov:2010yq,Borsanyi:2012zv,Budapest-Marseille-Wuppertal:2013vij,Boyle:2015exm,Cossu:2016eqs,Aoki:2017paw} and even $2+1+1$ quark flavors~\cite{Cichy:2013gja,Alexandrou:2017bzk} (see also Refs.~\cite{Narayanan:2004cp,Bali:2013kia,Hernandez:2019qed,Perez:2020vbn,DeGrand:2023hzz,Bonanno:2023ypf} for numerical calculations of QCD LECs in the large-$N_c$ limit). Predictions have been continuously refined over the time, finding overall a remarkably good agreement at the percent level among determinations obtained with a variety of lattice discretizations and numerical methods~\cite{FlavourLatticeAveragingGroupFLAG:2021npn}.

This paper can be placed within this context, as it deals with the problem of accurately determining QCD LECs on the lattice. In particular, we present an extensive calculation of the $\SU(2)$ chiral condensate in $2+1$ QCD using staggered quarks, combining 3 different strategies, and performing controlled continuum and chiral extrapolations. To this end, we will exploit results obtained from simulations performed at 4 values of the lattice spacing, ranging from $\sim 0.15$ to $\sim 0.07$ fm, and considering 4 different lines of constant physics, corresponding to
4 values of the pseudo-Goldstone pion mass ranging from $m_\pi^{\phys}$ to $3 m_\pi^{\phys}$, with the strange quark mass kept fixed at the physical point for all ensembles.

Concerning the numerical strategies pursued in this work, the main one will rely on the Giusti--L\"uscher method to extract the chiral condensate from the mode number of the Dirac operator~\cite{Giusti:2008vb,Luscher:2010ik}, which is implemented and applied to the case of staggered quarks for the first time in this paper. Another method consists in extracting the chiral condensate from the quark mass dependence of the topological susceptibility. Finally, we will also check consistency with the more standard and well-established method relying on the quark mass dependence of the pion mass, which, in the case of staggered quarks, has been already employed in Ref.~\cite{Borsanyi:2012zv} to determine $\Sigma$.

This manuscript is organized as follows. In Sec.~\ref{sec:latset} we present our numerical setup. In Sec.~\ref{sec:results} we present our continuum and chiral extrapolated results for the chiral condensate from the 3 different and complementary strategies outlined above. Finally, in Sec.~\ref{sec:conclu} we draw our conclusions.

\section{Numerical setup}\label{sec:latset}

In the first part of this Section we illustrate the discretization of $N_f = 2+1$ QCD on different Lines of Constant Physics (LCPs) adopted in our study. Then we move to a detailed discussion of the three numerical strategies exploited for the determination of the chiral condensate.

\subsection{Lattice action and determination of LCPs}

We perform simulations of $N_f=2+1$ QCD discretized on $N_s^4$ hypercubic lattices with two degenerate light quark flavors $u=d=l$, with bare mass $m_u=m_d\equiv m_l$, and a strange quark flavor $s$ with bare mass $m_s$. We adopt rooted stout staggered discretization of the Dirac operator:
\begin{gather}
\mathcal{M}^{(\stag)}_f[U] \equiv D_{\stag}[U^{(2)}] + am_f\,,\nonumber\\
\label{eq:stag_operator}
D_{\stag}[U^{(2)}] = \sum_{\mu=1}^{4}\eta_{\mu}(x)\left[ U^{(2)}_{\mu}(x) \delta_{x,y-\hat{\mu}} - {U_{\mu}^{(2)}}^{\dagger}(x-a\hat{\mu}) \delta_{x,y+\hat{\mu}} \right]\ ,\\ 
\eta_{\mu}(x) = (-1)^{x_1+\,\dots\,+x_{\mu-1}}\ , \nonumber
\end{gather}
where $U^{(2)}_\mu(x)$ are the gauge links after two steps of isotropic stout smearing with $\rho_{\stout}=0.15$~\cite{Morningstar:2003gk}. The gauge sector is instead discretized by using the tree-level Symanzik-improved action:
\beq
S_{\YM}^{(L)}[U] = - \frac{\beta}{3} \sum_{x, \mu \ne \nu} \left\{ \frac{5}{6}\Re\Tr\left[\Pi_{\mu\nu}^{(1\times1)}(x)\right] 
- \frac{1}{12}\Re\Tr\left[\Pi_{\mu\nu}^{(1\times2)}(x)\right]\right\}\ ,
\eeq
where $\Pi_{\mu\nu}^{(n\times m)}(x)$ stands for the product of non-stouted links along $n\times m$ rectangular paths, starting in the point $x$ and extending in the $(\mu,\nu)$ plane.
In the end, the partition function on the lattice can be written in the following way:
\beq\label{eq:partition_function}
Z_{\LQCD}=\int [dU] e^{-S^{(L)}_\YM[U]} \det\left\{\mathcal{M}^{(\stag)}_{l}[U]\right\}^{\frac{1}{2}} 
\det\left\{\mathcal{M}^{(\stag)}_{s}[U]\right\}^{\frac{1}{4}}\ .
\eeq
The sampling of the functional integral was performed by means of the standard Rational Hybrid Monte Carlo (RHMC) algorithm~\cite{Clark:2006fx,Clark:2006wp}.

In order to compute the $\SU(2)$ chiral condensate, one has to take the chiral limit $m_l\rightarrow0$ at fixed physical strange quark mass. To this end, we need to determine LCPs corresponding to different values of $m_l$ and fixed $m_s = m_s^{\phys}$.

Our starting point is the LCP with $m_l = m_l^{\phys}$ (i.e., corresponding to the physical value of the pion mass $M_\pi^{\phys}\simeq135~\mathrm{MeV}$) and with $R \equiv m_l/m_s = R^{\phys} \simeq 1/28.15$, that was determined in Refs.~\cite{Aoki:2009sc, Borsanyi:2010cj, Borsanyi:2013bia}, using the same discretization adopted here\footnote{The physical light-to-strange quark mass ratio $R^{\phys} \simeq 1/28.15$ was determined in~\cite{Aoki:2009sc, Borsanyi:2010cj, Borsanyi:2013bia} with a $\sim 0.4\%$ precision. Since we verified that, in our chiral extrapolations, the error on this quantity is completely negligible, being the error on the chiral condensate at least 3 times larger in the best case, in the following calculations we will simply neglect the errors on $R=m_l/m_s$.}.

To obtain LCPs corresponding to different values of the pion mass at fixed physical strange quark mass, we assumed that there is a window within which the light quark mass $m_l$ can be varied, at fixed $\beta$ and $m_s$, without having a significant impact on the value of the lattice spacing. Although it is well known from arguments based on the hopping-parameter expansion that the effective $\beta$ changes as we approach the quenched limit, our assumption is not unreasonable provided that we do not go too far from the physical point. Thus, we defined our LCPs with heavier-than-physical pions and physical strange quark as follows: after selecting a point $(\beta^{\phys},m_l^{\phys}, m_s^{\phys})$ along the physical LCP, we changed the bare light mass $m_l$ keeping $\beta = \beta^{\phys}$ and $m_s = m_s^{\phys}$ fixed, where the superscript ``$\phys$'' stands for bare parameters drawn from the physical LCP. We here note that the same procedure, with the same lattice discretization used here, was followed in Ref.~\cite{Borsanyi:2012zv} to take the chiral limit.

In order to verify that our assumptions are correct, we explicitly checked that the value of the lattice spacing did not change within our typical uncertainties. We fixed the scale by using the quantity $w_0$, based on the gradient flow~\cite{BMW:2012hcm}. Such quantity was assumed independent of the pion mass, and the same value was used to obtain $a$ for all the ensembles considered in this work (cf.~also Ref.~\cite{BMW:2012hcm} on this point). Using this prescription, we observe that the lattice spacing changes at most by $\sim 2\%$ among ensembles with different values of $m_l$, i.e., within the quoted uncertainties for $a$, thus confirming the validity of our procedure to fix the LCPs with unphysical pions.

We performed simulations for $m_l=4,6,9~m_l^{\phys}$ with $\beta = \beta^{\phys}$ and $m_s = m_s^{\phys}$. Concerning simulations at the physical point, instead, we used some of the gauge ensembles that were previously generated for our paper~\cite{Athenodorou:2022aay}. Finally, concerning the lattice volume, the size of the box was always chosen in order to stay within the range $\ell \equiv a N_s = 2.7$ -- $3.1$ fm, corresponding to $M_\pi \ell \gtrsim 2$. As we will discuss in Sec.~\ref{sec:BanksCasher_lattice}, this is largely sufficient to contain finite size effects affecting the computation of the chiral condensate. A full summary of our simulation parameters is shown in Tab.~\ref{tab:simulation_parameters}.

\begin{table}[!htb]
\begin{center}
\begin{tabular}{ |c|c|c|c|c|c|c|}
\hline
$R\equiv m_l / m_s$ & $m_l / m_l^{\phys}$ & $\beta$ & $N_s$ & $a m_l$ & $a m_s$ & $a$~[fm] \\
\hline
\hline
\multirow{4}{*}{\makecell{1/28.15\\$\simeq$ 0.0355}} & \multirow{4}{*}{1} & 3.750 & 24 & 0.0018 & 0.0503 & 0.1249  \\
&& 3.850 & 32 & 0.0014 & 0.0394 & 0.0989  \\
&& 3.938 & 32 & 0.0012 & 0.0330 & 0.0824  \\
&& 4.020 & 40 & 0.0010 & 0.0281 & 0.0707  \\

\hline	
\hline		
\multirow{4}{*}{\makecell{4/28.15\\$\simeq$ 0.1421}} & \multirow{4}{*}{4} & 3.678 & 20 & 0.0088 & 0.0621 & 0.1515   \\
&& 3.750 & 24 & 0.0072 & 0.0503 & 0.1265 \\
&& 3.868 & 32 & 0.0054 & 0.0379 & 0.0964  \\
&& 3.988 & 40 & 0.0042 & 0.0299 & 0.0758  \\
\hline
\hline		
\multirow{4}{*}{\makecell{6/28.15\\$\simeq$ 0.2131}} & \multirow{4}{*}{6} & 3.678 & 20 & 0.0132 & 0.0621 & 0.1532   \\
&& 3.750 & 24 & 0.0107 & 0.0503 & 0.1278  \\
&& 3.868 & 32 & 0.0081 & 0.0379 & 0.0976  \\
&& 3.988 & 40 & 0.0064 & 0.0299 & 0.0764  \\
\hline
\hline
\multirow{4}{*}{\makecell{9/28.15\\$\simeq$ 0.3197}} & \multirow{4}{*}{9} & 3.678 & 20 & 0.0199 & 0.0621 & 0.1556  \\
&& 3.750 & 24 & 0.0161 & 0.0503 & 0.1297  \\
&& 3.868 & 32 & 0.0121 & 0.0379 & 0.0989  \\
&& 3.988 & 40 & 0.0095 & 0.0299 & 0.0768  \\
\hline
\end{tabular}
\end{center}
\caption{Summary of simulation parameters used in this work. The bare parameters and lattice spacings of the points with $m_l=m_l^{\phys}$ have been fixed according to the LCP determined in Refs.~\cite{Aoki:2009sc, Borsanyi:2010cj, Borsanyi:2013bia}, with $M_\pi = M_\pi^{\phys} =135~\mathrm{MeV}$ and $R = m_l/m_s= R^{\phys} = 1/28.15$. Concerning points with $m_l/m_l^{\phys}\neq1$, the lattice spacings have been determined with the $w_0$ scale setting approach~\cite{BMW:2012hcm} with a $\sim 2\%$ error, and their value in $\mathrm{fm}$ units was obtained assuming $w_0 = 0.1757(12)~\mathrm{fm}$~\cite{BMW:2012hcm} for all ensembles, independently of the pion mass.}
\label{tab:simulation_parameters}
\end{table}

\subsection{The Banks--Casher relation on the lattice}\label{sec:BanksCasher_lattice}

The properties of the low-lying spectrum of the massless Dirac operator $\slashed{D}$ encode information about the chiral condensate via the Banks--Casher relation~\cite{Banks:1979yr}:
\beq\label{eq:banks_casher}
\lim_{\lambda\to 0}\lim_{m\rightarrow0}\lim_{V\rightarrow\infty}\rho(\lambda,m) = \frac{\Sigma}{\pi},
\eeq
where $\rho(\lambda,m)$ is the spectral density of the eigenvalues $i\lambda$ of $\slashed{D}$,
\beq\label{eq:spectral_density_def}
\rho(\lambda,m)=\frac{1}{V}\sum_{k} \langle \delta(\lambda-\lambda_k)\rangle.
\eeq

On the lattice, in order to extract the chiral condensate from the low-lying spectrum of the lattice Dirac operator, a more convenient quantity to work with is the \textit{mode number}~\cite{Giusti:2008vb}:
\beq\label{eq:mode_number_def}
\braket{\nu(M)}=V\int_{-M}^{M}\rho(\lambda,m) d \lambda,
\eeq
where $V$ is the $4d$ space-time volume, and $\braket{\nu(M)}$ stands for the mean number of eigenmodes of $\slashed{D}$ whose absolute value lies below the threshold $M$. This quantity contains the same physical information with respect to the spectral density; as a matter of fact, in the chiral limit and sufficiently close to the origin, we expect from the Banks--Casher relation:
\beq\label{eq:mode_num_sigma}
\braket{\nu(M)} = \frac{2}\pi V\Sigma M,
\eeq
plus higher-order terms in $M$. The method proposed by Giusti and L\"uscher~\cite{Giusti:2008vb} to extract the chiral condensate is based on Eq.~\eqref{eq:mode_num_sigma}: it consists in computing numerically the mode number, then performing a linear fit of $\braket{\nu}$ as a function of $M$ in a region close enough to the origin. The slope of such linear fit is the condensate, modulo overall factors.

The mode number can be determined quite accurately on the lattice, either by means of noisy estimators (as done in Ref.~\cite{Giusti:2008vb}) or by directly computing the first lowest eigenvalues of the discretized Dirac operator, which is the strategy followed in this paper. In the case of the massless staggered Dirac operator $D_\stag$, the spectrum is made of purely imaginary numbers, which appear in complex conjugate pairs. Therefore, we solved the following eigenproblem numerically, using the \texttt{ARPACK} libraries and computing the first few hundred eigenvalues and eigenvectors:
\beq\label{eq:eigenproblem}
i D_\stag u_\lambda = \lambda u_\lambda, \qquad \lambda \in \mathbb{R},
\eeq
where $D_\stag$ is the very same operator used for sea quarks. Then, we simply computed:
\beq
\braket{\nu_\stag(M)} = 2\braket{\# \lambda \text{ with } 0<\lambda<M},
\eeq
where $\nu_\stag$ stands for the number of staggered modes falling below $M$. Since the spectrum of $D_\stag$ becomes four-fold degenerate in the continuum limit, due to the taste degeneracy, in the continuum limit we expect the counting of staggered modes to become equal to $n_t$ times the counting of modes of the Dirac operator for a single flavor, where $n_t=2^{d/2}=4$ denotes the number of different staggered tastes. Therefore, in order to recover the proper continuum limit for the mode number, and thus for the chiral condensate, we have to take into account such mode over-counting as follows:
\beq
\braket{\nu_\stag(M)} = n_t \braket{\nu(M)}.
\eeq
Thus, Eq.~\eqref{eq:mode_num_sigma} with staggered quarks is modified as follows:
\beq\label{eq:mode_num_sigma_stag}
\braket{\nu_\stag(M)} = n_t\frac{2}\pi V\Sigma M.
\eeq

Let us now discuss how we renormalize Eq.~\eqref{eq:mode_num_sigma_stag} in our setup. The spectral threshold $M$ renormalizes as a quark mass~\cite{Giusti:2008vb,Bonanno:2019xhg}, which in a staggered discretization renormalizes only 
multiplicatively. Therefore, the ratio of $M$ to any quark mass is a Renormalization Group (RG) invariant quantity. Since in our setup the strange mass is always kept at the physical point and the light mass is varied, it is natural to use the former to renormalize $M$. Using the fact that the mode number is an automatically RG-invariant quantity~\cite{Giusti:2008vb}, it is clear that it is sufficient to multiply and divide for $m_s$ in Eq.~\eqref{eq:mode_num_sigma_stag} to obtain a fully renormalized relation that can be used in our numerical calculations:
\beq\label{eq:mode_num_sigma_stag_RENORM}
\braket{\nu_\stag(M/m_s)} = n_t \frac{2}{\pi} V \left(\Sigma m_s\right) \left(\frac{M}{m_s}\right).
\eeq
Therefore, by performing a linear fit of the staggered mode number $\braket{\nu_\stag}$ as a function of the RG-invariant ratio $M/m_s$, we can extract the RG-invariant quantity $\Sigma m_s$ according to Eq.~\eqref{eq:mode_num_sigma_stag_RENORM}.

As a final remark, we want to discuss finite-mass and finite volume effects. As for the former, we stress that the quantity extracted from the linear best fit of the mode number can be interpreted as a physical quark condensate only in the chiral limit $m_l\to 0$. At finite light sea quark mass, the quantity we extract will be actually an ``effective'' mass-dependent condensate, needing extrapolation towards the chiral limit. The mass dependence of the effective condensate has been worked out in~\cite{Giusti:2008vb} using ChPT and, at LO in the quark mass, is expected to be linear in $m_l$.

Concerning finite volume effects, in Ref.~\cite{Giusti:2008vb} the authors have also worked out the approach to the thermodynamic limit of the effective condensate within ChPT, finding that finite volume corrections are not exponentially suppressed in $M_\pi \ell$ but in $M_0 \ell$, where the relevant scale is given by:
\beq\label{eq:fss_chircond_modenumber}
M_0^2 = \frac{2 \Sigma M}{F^2_\pi} = M_\pi^2 \frac{1}{R}\frac{M}{m_s},
\eeq
with $M$ the same scale appearing in Eq.~\eqref{eq:mode_number_def} and $\ell\equiv a N_s$. As we will show in the following, the sizes of our lattices, which satisfy $M_\pi \ell \gtrsim 2$, and the value of $M/m_s$ used in this work, are sufficient to contain finite size effects, as they correspond to $M_0 \ell \gtrsim 4$.

\subsection{The quark mass dependence of the pion mass}

There are different and complementary ways of computing the chiral condensate from the lattice other than the mode number. The most straightforward one relies on the GMOR relation in Eq.~\eqref{eq:GMOR_nondegenerate}, which in the case of 2 degenerate light flavors becomes:
\beq\label{eq:GMOR}
M^2_\pi = \frac{2 \Sigma}{F^2_\pi} m_l.
\eeq
From the point of view of lattice calculations, numerical determinations of the pion mass can be fitted to Eq.~\eqref{eq:GMOR} as a function of $m_l$ to extract the ratio of LECs $\Sigma/F_\pi^2$, from which we can extract $\Sigma$ once $F_\pi$ has been taken out.

Both the pion mass and the pion decay constant can be extracted from the correlator of the appropriate lattice staggered interpolating operator of the taste pseudoscalar pion $P(t) = \sum_{\vec{x}} P(t,\vec{x})$~\cite{MILC:2009mpl}. As a matter of fact, for sufficiently large time separations, the correlator is expected to be described by a single exponential of the type:
\beq\label{eq:fit_exp_pion_tcorr}
C_\pi(t) = \braket{P(t)P(0)} \underset{t \to \infty}{\sim} A_\pi\left[ e^{-M_\pi t} + e^{-M_\pi(\ell- t)} \right],
\eeq
where $M_\pi$ is the pion mass, and where the pre-factor $A_\pi$ is the matrix element
\beq
A_\pi = \vert\bra{\Omega} \overline{u} \,\epsilon\, d \ket{\pi(\vec{p}=\vec{0})}\vert^2/M_\pi,
\eeq
with $\epsilon$ playing the role of $\gamma_5 \otimes \xi_5$ in spinor and taste space~\cite{MILC:2009mpl}. Recalling the definition of $F_\pi$ in Eq.~\eqref{eq:Fpi_def} and using the PCAC relation, one quickly arrives to the following relation~\cite{Borsanyi:2012zv}:
\beq\label{eq:f_pi_from_pion_fit}
F_\pi = m_l \sqrt{\frac{A_\pi}{M_\pi^3}} = m_s R \sqrt{\frac{A_\pi}{M_\pi^3}}.
\eeq

\subsection{The quark mass dependence of the topological susceptibility}\label{sec:topsusc}

Another strategy to obtain the chiral condensate is to study the behavior of the topological susceptibility (here $G_{\mu\nu}$ denotes the gauge field strength tensor)
\beq
\chi=\frac{\braket{Q^2}}{V}, \qquad Q = \frac{1}{32\pi^2} \varepsilon_{\mu\nu\rho\sigma} \int \Tr\left\{G_{\mu\nu}(x)G_{\rho\sigma}(x)\right\} d^4x,
\eeq
as a function of the light quark mass. As a matter of fact, ChPT with 2 degenerate light flavors predicts at LO:
\beq\label{eq:ChPT_topsusc}
\chi = \frac{1}{2} \Sigma m_l,
\eeq
where the vanishing of $\chi$ in the chiral limit is a well known hallmark of the Index theorem. Thus, the chiral condensate can also be inferred from the slope of the topological susceptibility as a function of $m_l$ (see, e.g., Ref.~\cite{Alexandrou:2017bzk}, where such strategy was used adopting Twisted Mass Wilson fermions).

Several equivalent definitions of the topological charge can be taken on the lattice, differing among themselves for lattice artifacts, but all agreeing in the continuum limit. They can be broadly divided into gluonic definitions, based on straightforward discretizations of $G\widetilde{G}$ computed on smoothened configurations~\cite{Berg:1981nw,Iwasaki:1983bv,Itoh:1984pr,Teper:1985rb,Ilgenfritz:1985dz,Campostrini:1989dh,Alles:2000sc,Bonati:2014tqa, Alexandrou:2015yba,Luscher:2009eq, Luscher:2010iy,DelDebbio:2002xa,Bonati:2015sqt}, and fermionic definitions, based instead on the Index theorem.

As it has been extensively reported in the recent literature, gluonic definitions of the topological susceptibility are typically affected by large corrections to the continuum limit~\cite{Bonati:2015vqz,Petreczky:2016vrs,Borsanyi:2016ksw,Frison:2016vuc,Alexandrou:2017bzk,Bonati:2018blm,Athenodorou:2022aay}. On the other hand, definitions based on spectral projectors of the lowest-lying modes of the discretized Dirac operator have been shown numerically to suffer for much smaller lattice artifacts, allowing to control systematic errors on the continuum extrapolation more easily~\cite{Giusti:2008vb,Luscher:2010ik,Cichy:2015jra,
Alexandrou:2017bzk, Athenodorou:2022aay}.

In this work we will consider a fermionic definition of the topological charge, based on spectral projectors~\cite{Giusti:2008vb,Luscher:2010ik,Cichy:2015jra,
Alexandrou:2017bzk} on the lowest-lying modes of the staggered operator. Such definition, worked out and probed in the quenched case in Ref.~\cite{Bonanno:2019xhg}, has been applied both at zero and at finite temperature in Ref.~\cite{Athenodorou:2022aay}, where it has been shown to yield agreeing results with the standard gluonic definition in the continuum limit, but suffering for much smaller lattice artifacts.

The idea is to define a bare lattice topological charge as
\beq\label{eq:def_QSP}
Q_\SP^{(0)} = \frac{1}{n_t} \sum_{\vert \lambda \vert \le M} r_\lambda, \qquad r_\lambda = u_\lambda^\dagger \Gamma_5 u_\lambda,
\eeq
with $\Gamma_5$ the staggered definition of $\gamma_5$. Here, the pseudo-chiralities $r_\lambda$, defined from the very same eigenvectors $u_\lambda$ of the eigenproblem in Eq.~\eqref{eq:eigenproblem}, are real numbers with absolute value between 0 and 1. In the continuum limit, instead, we expect $r_{\lambda\,=\,0}=\pm 1$ and $r_{\lambda\,\neq\,0}=0$, thus Eq.~\eqref{eq:def_QSP} will simply reduce to the Index theorem for any value of $M$, which at this level is a free parameter that can be thought of as an intrinsic UV cut-off of the spectral definition, similar to the UV cut-off introduced by smoothing when considering gluonic definitions. Note again the presence of a factor of $1/n_t$ to take into account the mode over-counting due to taste degeneracy.

The bare definition of the topological charge in Eq.~\eqref{eq:def_QSP} can be proven to renormalize only multiplicatively~\cite{Giusti:2008vb,Bonanno:2019xhg}:
\beq\label{eq:SP_charge_RENORM}
Q_{\SP} &=& Z_{\SP} Q_{\SP}^{(0)},\\
\label{eq:SP_renorm_const}
Z_{\SP} &=& 
\sqrt{\frac{\braket{\Tr\left\{\mathbb{P}_M\right\}}}{\braket{\Tr\left\{\Gamma_5 \mathbb{P}_M \Gamma_5 \mathbb{P}_M\right\}}}}\ ,
\eeq
where we have introduced the spectral projector
\beq\label{eq:projectors}
\mathbb{P}_M \equiv \sum_{\vert \lambda \vert \le M} u_{\lambda} u_{\lambda}^{\dagger}\ .
\eeq
Using Eq.~\eqref{eq:projectors}, the bare definition in Eq.~\eqref{eq:def_QSP} can be rewritten as:
\beq
Q_\SP^{(0)} = \Tr\left\{\Gamma_5 \mathbb{P}_M \right\}.
\eeq

The traces appearing in Eq.~\eqref{eq:SP_renorm_const} can be easily computed from the eigenvalues and eigenvectors of $D_\stag.$ As a matter of fact, $\braket{\Tr\left\{\mathbb{P}_M\right\}}$ is simply the mode number $\braket{\nu_\stag(M)}$; the denominator, instead, can be computed from the following spectral sum:
\beq
\braket{\Tr\left\{\Gamma_5 \mathbb{P}_M \Gamma_5 \mathbb{P}_M\right\}} = \sum_{\vert \lambda \vert \le M} \sum_{\vert \lambda^\prime \vert \le M} \vert u_{\lambda^\prime}^\dagger \Gamma_5 u_\lambda \vert^2.
\eeq
The final spectral expression of the topological susceptibility can then be written as:
\beq\label{eq:chi_SP_final}
\chi_{\SP} = \frac{\left\langle Q_{\SP}^2\right\rangle}{V} =  
\frac{1}{n_t^2}\frac{\braket{\Tr\{\mathbb{P}_M\}}}{\braket{\Tr\{\Gamma_5 \mathbb{P}_M \Gamma_5 \mathbb{P}_M \}}} \frac{\braket{\Tr\{\Gamma_5 \mathbb{P}_M\}^2}}{V}\ .
\eeq

As a final remark, we briefly discuss the role of $M$. As we have already stressed, continuum results should not depend on $M$, as in that limit only zero-modes will contribute to the topological charge. Thus, we only expect lattice artifacts to depend on the choice of this threshold. In Refs.~\cite{Alexandrou:2017bzk,Athenodorou:2022aay} it was actually shown that this is the case, and that with a suitable choice of $M$ it is possible to reduce dramatically lattice artifacts affecting the gluonic definition of the topological susceptibility. Following the lines of Ref.~\cite{Athenodorou:2022aay}, we will thus compute the continuum limit of $\chi_\SP$ for several values of $M$, in order to ensure that the choice of this quantity does not introduce any source of systematic errors.

\section{Numerical results}\label{sec:results}

In this section we will present our numerical results for the chiral condensate. We will present three different calculations, relying respectively on the mode number, the pion mass and the topological susceptibility. In all cases, we will perform a controlled continuum limit at fixed pion mass, followed by a controlled chiral extrapolation. Finally, we will also perform a global fit of our data imposing that they are all described by the same value of the condensate.

\subsection{Chiral condensate from the mode number}\label{sec:sigma_from_mode_number}

\begin{figure}[!htb]
\centering
\includegraphics[scale=0.6]{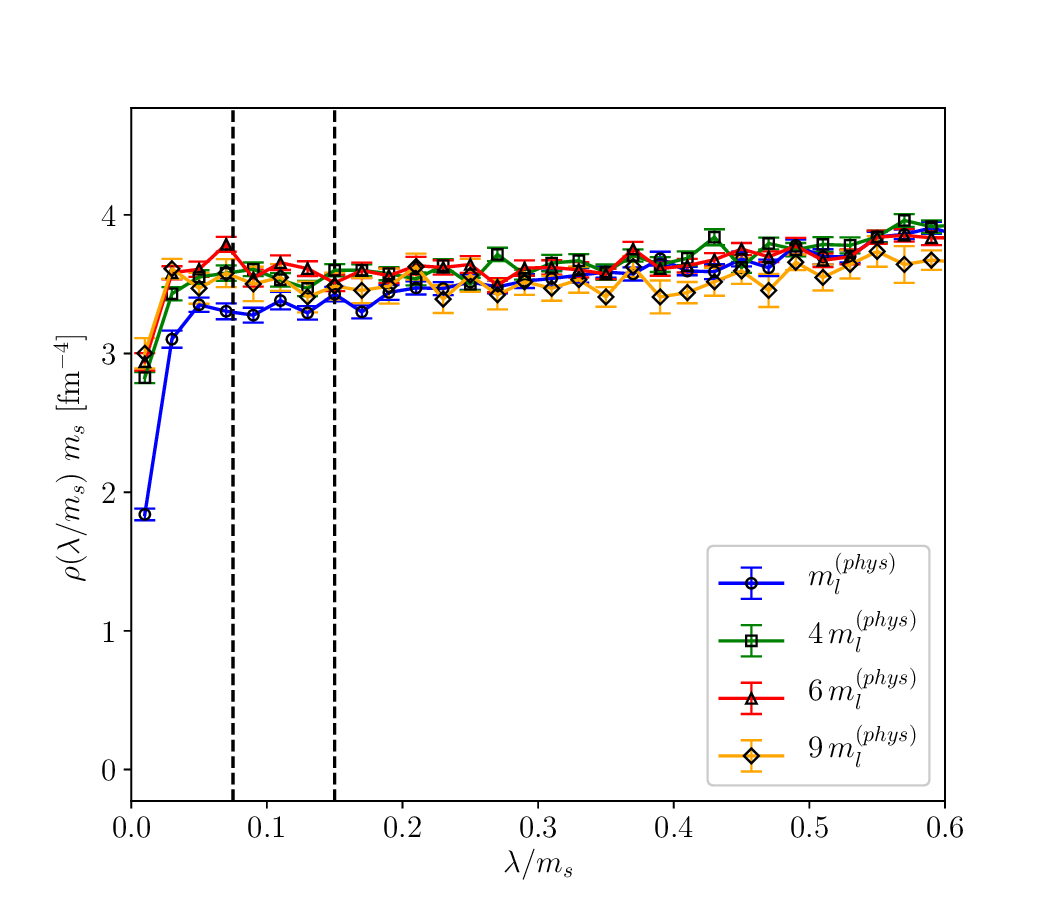}
\caption{Behavior of the RG-invariant spectral density $m_s\rho(\lambda/m_s)$ as a function of the RG-invariant ratio $\lambda/m_s$ for the lattices with the finest lattice spacing, and for all explored values of $m_l$. The range between the two dashed vertical lines, $\lambda/m_s\in[0.075,0.15]$, is the one chosen to perform the linear best fit of the mode number.}
\label{fig:spectral_density}
\end{figure}

In this Section we will discuss our determination of the chiral condensate from the mode number. First of all, in order to determine a reasonable fit range for the mode number where higher-order corrections in $M$ are negligibile, we have looked for a common plateau region for the spectral densities determined for the finest lattice spacings at the various quark masses. Such determinations are shown in Fig.~\ref{fig:spectral_density}. From our results it is clear that in the range $\lambda /m_s \in [0.075,0.15]$ the spectral density is fairly constant for all ensembles, thus it is reasonable to assume that higher-order terms in $\lambda$ can be neglected in this interval. Therefore, we will use this range for the linear best fit of the mode number using the Giusti--L\"uscher method.

\begin{figure}[!htb]
\centering
\includegraphics[scale=0.42]{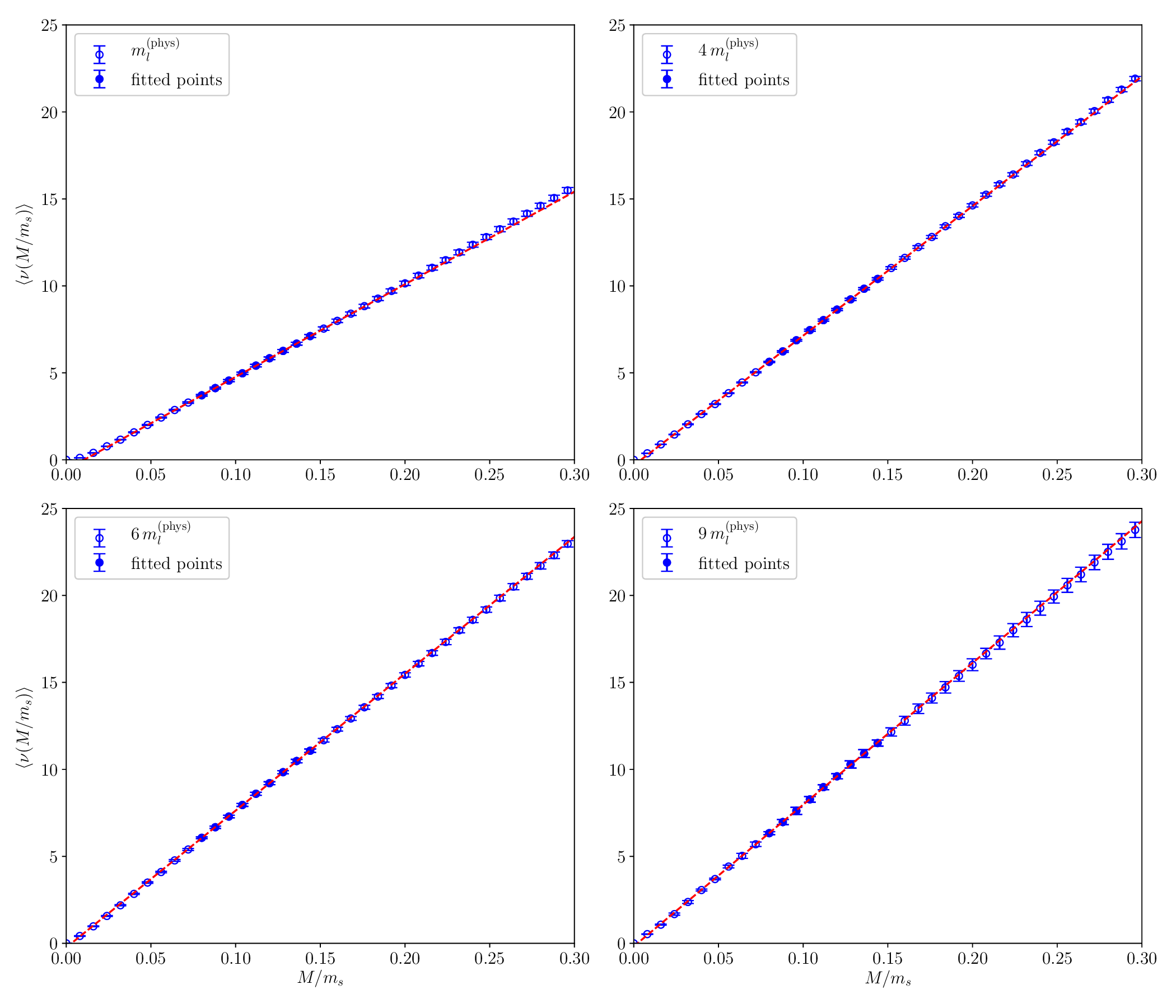}
\caption{Linear best fit of the physical mode number $\braket{\nu} = \braket{\nu_\stag}/4$ as a function of $M/m_s$ for the finest lattice spacing for each value of $m_l$. Filled points in the range $M/m_s\in[0.075,0.15]$ are those included in the best fits, depicted as dashed lines.}
\label{fig:fit_mode_number}
\end{figure}

Within the range earlier determined, we observe from Fig.~\ref{fig:fit_mode_number} that, as expected, the mode number can be reliably described by a linear rise in $M/m_s$. Performing the following best fit, where $s$ denotes the slope of the mode number in the middle point of the fit range,
\beq
\braket{\nu_\stag(M/m_s)} = s \frac{M}{m_s},
\eeq
we obtain the effective chiral condensate as:
\beq
\Sigma m_s = \frac{\pi}{2} \frac{s}{n_t V}, \qquad V = (aN_s)^4.
\eeq
In order to take out the strange quark mass in the latter expression for $\Sigma m_s = \Sigma_\R m_s^{(\R)}$, we use the following value for the renormalized mass:
\beq\label{eq_ms_phys}
m_s^{(\R)}=92.4(1.5)~\mathrm{MeV}.
\eeq
This value was obtained in the $\overline{\mathrm{MS}}$ renormalization scheme at the conventional renormalization point $\mu=2$ GeV from a $2+1$ lattice QCD calculation involving staggered fermions in Ref.~\cite{Davies:2009ih}. After removing the factor of $m_s^{(\R)}$ via the renormalized mass above, we are thus left with the renormalized chiral condensate $\Sigma_\R$, which of course has to be understood as expressed in the $\overline{\mathrm{MS}}$ scheme at $\mu=2$ GeV as well. Our determinations of $\Sigma_\R$ as a function of the lattice spacing $a$ and of the light quark mass $m_l$ are collected in Tab.~\ref{tab:eff_chiral_condensate_values}.

\FloatBarrier

\begin{table}[!t]
\begin{center}
\begin{tabular}{ |c|c|c|c|}
\hline
$R\equiv m_l / m_s$ & $m_l / m_l^{\phys}$    & $a$~[fm] & $\Sigma^{1/3}_{\mathrm{R}}$~[MeV] \\
\hline
\hline
\multirow{5}{*}{\makecell{1/28.15\\$\simeq$ 0.0355}} & \multirow{5}{*}{1} & 0.1249 &282.0(1.6) \\
&& 0.0989 & 282.1(1.6) \\
&& 0.0824 & 282.8(1.7) \\
&& 0.0707 & 277.9(1.6) \\
&& 0      & $275.4(3.5)_{\mathrm{stat}}(3.7)_{\mathrm{sys}}$ \\

\hline	
\hline		
\multirow{5}{*}{\makecell{4/28.15\\$\simeq$ 0.1421}} & \multirow{5}{*}{4} & 0.1515 & 284.3(1.5) \\
&& 0.1265 & 283.0(1.5)\\
&& 0.0964 & 283.3(1.5)\\
&& 0.0758 & 283.7(1.6)\\
&& 0      & $284.1(2.4)_{\mathrm{stat}}(0.4)_{\mathrm{sys}}$\\
\hline
\hline		
\multirow{5}{*}{\makecell{6/28.15\\$\simeq$ 0.2131}} & \multirow{5}{*}{6} & 0.1532 & 283.2(1.5)\\
&& 0.1278 & 282.2(1.5)\\
&& 0.0976 & 283.3(1.6)\\
&& 0.0764 & 285.5(1.6)\\
&& 0      & $286.7(2.4)_{\mathrm{stat}}(1.6)_{\mathrm{sys}}$\\
\hline
\hline
\multirow{5}{*}{\makecell{9/28.15\\$\simeq$ 0.3197}} & \multirow{5}{*}{9} & 0.1556 & 282.6(1.5)\\
&& 0.1297 & 281.0(1.5)\\
&& 0.0989 & 283.4(1.6)\\
&& 0.0768 & 286.8(1.9)\\
&& 0      & $289.0(2.6)_{\mathrm{stat}}(3.4)_{\mathrm{sys}}$\\
\hline
\end{tabular}
\end{center}
\caption{Values of the cubic root of the effective chiral condensate obtained from the mode number for each value of $m_l$ and for each lattice spacing. The values of $\Sigma^{1/3}_{\mathrm{R}}$ at $a=0$ correspond to the extrapolations towards the continuum limit.}
\label{tab:eff_chiral_condensate_values}
\end{table}

\FloatBarrier

\begin{figure}[!htb]
\centering
\includegraphics[scale=0.51]{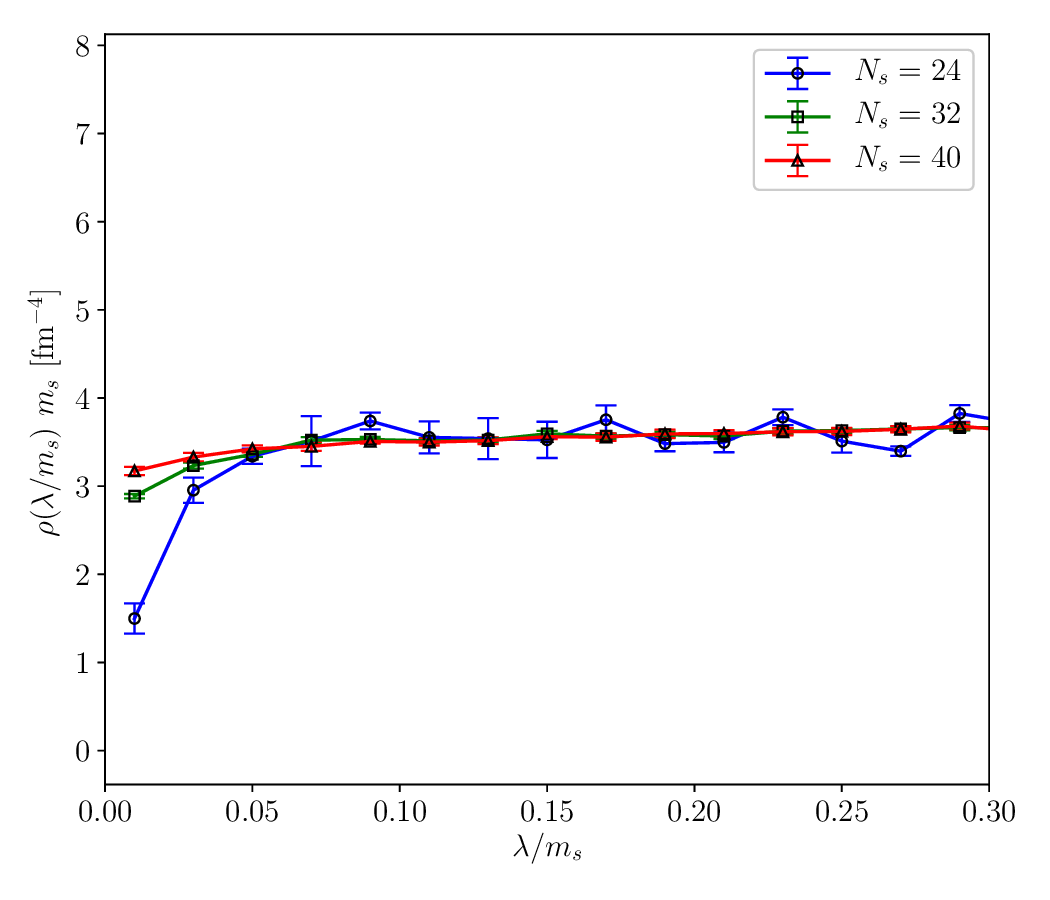}
\includegraphics[scale=0.51]{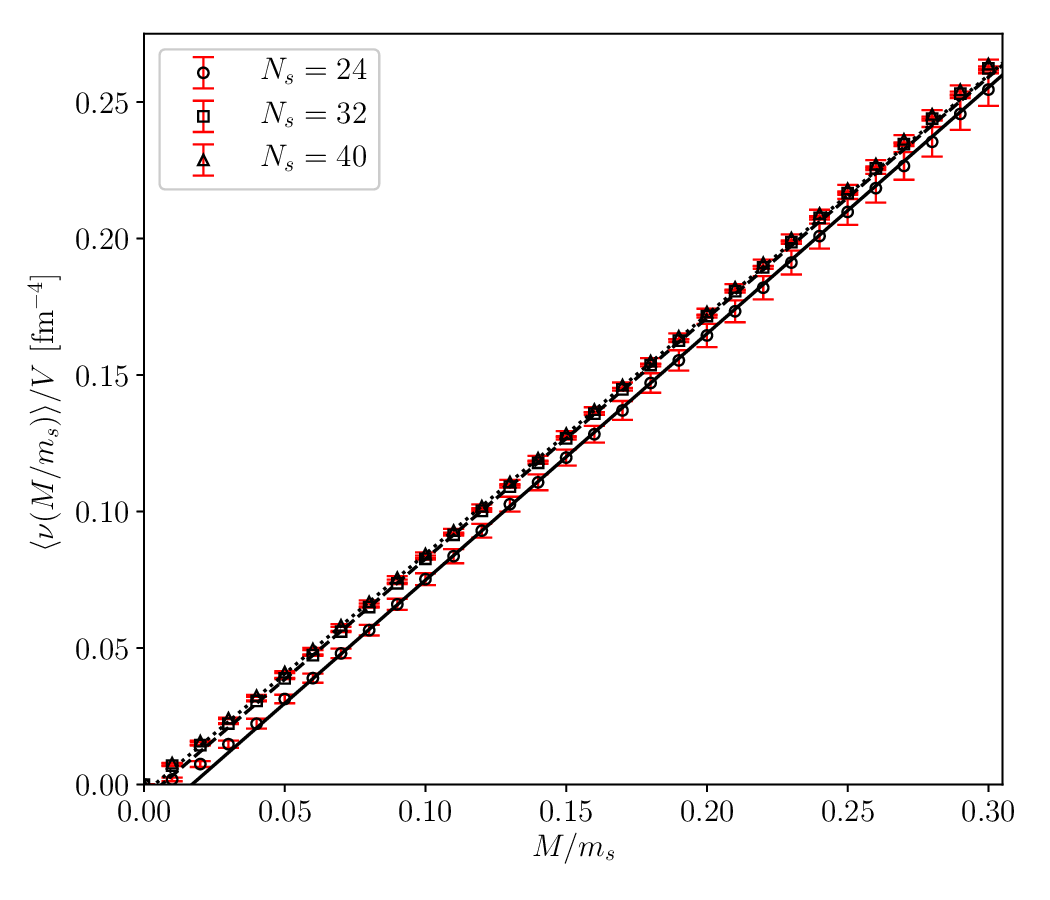}
\caption{Comparison of the spectral densities $m_s \rho(\lambda/m_s)$ (top panel) and of the mode number densities $\braket{\nu(M/m_s)}/V~[\mathrm{fm}^{-4}]$ (bottom panel) for 3 different lattices with sizes $N_s = 24,32,40$, corresponding to $M_0\ell \simeq 3,4,5$, for our point with $a=0.0964~\mathrm{fm}$, $\beta=3.868$ and $m_l = 4 m_l^{\phys}$.}
\label{fig:fss_modenumber}
\end{figure}

Before proceeding to discuss the continuum and chiral limits, let us make a remark about Finite Size Effects (FSEs). First of all, plugging $M_\pi=135$ MeV, $M/m_s = 0.15$ and $1/R=28.15$ in Eq.~\eqref{eq:fss_chircond_modenumber}, we obtain that the mass scale controlling FSEs affecting $\Sigma$ computed from the mode number is $M_0 \simeq 2 M_\pi$. Since we use lattices with $M_\pi \ell \gtrsim 2$, we have $M_0 \ell \gtrsim 4$, which is largely sufficient to maintain the magnitude of finite volume corrections below our typical statistical error, of the order of a few percent. This estimation is also in agreement with the prescription given in~\cite{Giusti:2008vb}, where the authors estimate that lattices with $\ell \gtrsim 2$ fm ($M_\pi \ell \gtrsim 1.4$) are sufficient to keep finite volume effects below the percent level. In any case, for a particular point of our ensembles we also considered two additional lattice volumes in order to directly check the absence of any dependence on the box size for the volumes adopted in this work, cf.~Fig.~\ref{fig:fss_modenumber}.

As it can be appreciated, varying the lattice size from $M_0 \ell \simeq 3$ to $M_0\ell \simeq 5$ only yields a significant difference for the spectral density in the two smallest bins, which are those that are expected to suffer more from FSEs, cf.~Fig.~\ref{fig:fss_modenumber} on the top. Above $M/m_s = 0.05$, instead, we observe no significant difference in the spectral densities. Since we are fitting the mode number in $[0.075,0.150]$, we thus expect no FSEs in the slopes of the mode numbers. As a matter of fact, the mode numbers for these 3 lattices have parallel slopes, and $\braket{\nu}$ for $M_0\ell \simeq 3$ only differs from the ones obtained at $M_0 \ell \simeq 4$ and $M_0 \ell \simeq 5$ for an overall shift (due to the FSEs affecting the lowest modes), cf.~Fig.~\ref{fig:fss_modenumber} on the bottom. In the end, we obtain $\Sigma_{\R}^{1/3} = 285(2), 283(2), 283(2)$~MeV, for, respectively, $M_0 \ell = 3, 4, 5$. Thus, we conclude that our determinations of the condensate from the mode number obtained on lattices with $M_0 \ell \simeq 4$ or larger are not affected by significant FSEs at the current level of precision.

\begin{figure}[!htb]
\centering
\includegraphics[scale=0.42]{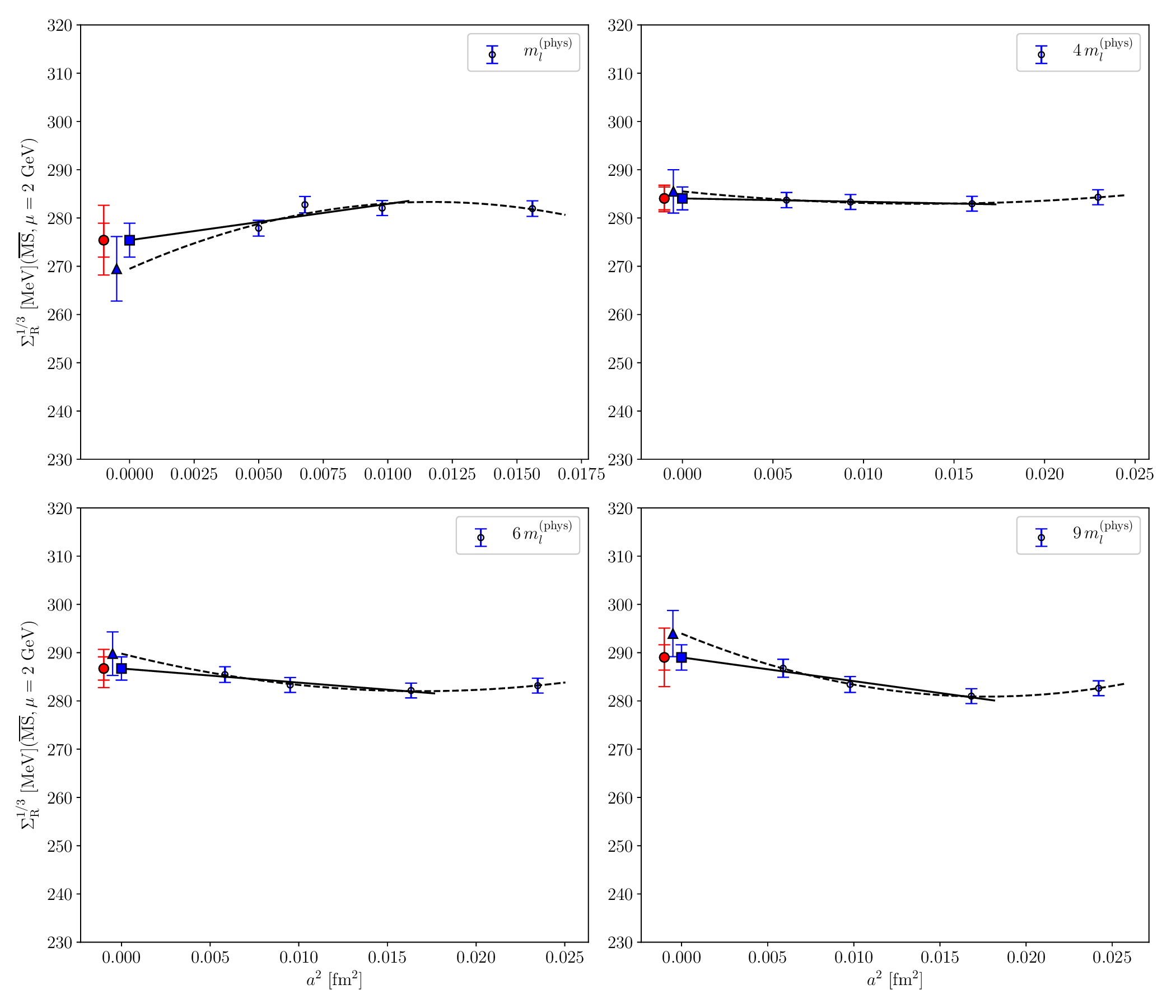}
\caption{Continuum limit extrapolation of $\Sigma_{\mathrm{R}}^{1/3}$ for each value of $m_l$ obtained from the mode number. Solid lines represent linear fits in $a^2$ to the 3 finest lattice spacings, dashed lines represent instead parabolic fits in $a^2$ to all of our data, including also the determinations at the coarsest lattice spacing. The square points in $a=0$ are the results from the linear fits, the triangular points the ones from the parabolic fits. Finally, the round points are the final estimations of $\Sigma_\R^{1/3}(R)$, with a double error bar referring to the statistical and to the sum of statistical and systematic uncertainties.}
\label{fig:cc_mode_number_continuum_limit}
\end{figure}

Once determinations of the effective chiral condensate have been obtained as a function of the lattice spacing $a$ and of the light quark mass $m_l$, we first take the continuum limit at fixed $m_l$, i.e., at fixed value of $R=m_l/m_s$, assuming standard $O(a^2)$ corrections: 
\beq
\Sigma_\R^{1/3}(a,R)=\Sigma_\R^{1/3}(R) + c_1(R)\,a^2+o(a^2).
\eeq
Continuum extrapolations are shown in Fig.~\ref{fig:cc_mode_number_continuum_limit}. As it can be observed, data for the three finest lattice spacings are perfectly described by linear corrections in $a^2$, and a parabolic best fit in $a^2$ of our data including also the coarsest lattice spacing gives agreeing results within the statistical errors. Therefore, we quote the results of the former fit as our values for the continuum limit.

In order to provide a more conservative estimate of the errors, we also assign to our continuum extrapolations a systematic error related to the small observed differences among the extrapolated values yielded by the two different fit ans\"atze employed. To do so, inspired by the procedure pursued in Refs.~\cite{ExtendedTwistedMassCollaborationETMC:2022sta,Bonanno:2023ljc,Bonanno:2023thi}, we first compute:
\beq\label{eq:discrepancy_error_bars}
\Delta = \frac{\left \vert \left [\Sigma^{1/3}_\R (R)\right]_{\mathrm{l3}} - \left [\Sigma^{1/3}_\R (R)\right]_{\mathrm{p4}} \right \vert}{\Delta_{\mathrm{stat}}\left [\Sigma^{1/3}_\R (R)\right ]_{\mathrm{p4}}},
\eeq
i.e., the difference between the central values of the continuum extrapolations obtained from the linear fit to the 3 finest points (denoted by ``l3") and from the parabolic fit to all points (denoted by ``p4"), weighted by the statistical error on the latter quantity. Finally, our systematic error is given by:
\beq\label{eq:systematic_error}
\Delta_{\mathrm{sys}}=\left \vert \left [\Sigma^{1/3}_\R (R)\right]_{\mathrm{l3}} - \left [\Sigma^{1/3}_\R (R)\right]_{\mathrm{p4}} \right \vert \operatorname{erf}\left(\frac{\Delta}{\sqrt{2}}\right)
\eeq
with $\operatorname{erf}(x)$ being the well-known \textit{error function},
\beq\label{eq:err_func}
\operatorname{erf}(x)=\frac{2}{\sqrt{\pi}}\int_0^x dt\ e^{-t^2}.
\eeq
In a few words, our systematic error in Eq.~\eqref{eq:systematic_error} is the difference between the continuum extrapolations obtained from the two employed fit ans\"atze, multiplied by the probability that this difference is due to a statistical fluctuation. All the continuum extrapolated values of $\Sigma_\R^{1/3}$ are reported in Tab.~\ref{tab:eff_chiral_condensate_values} as the determination for $a=0$, the first error is statistical, while the second is the systematic one, computed according to~\eqref{eq:systematic_error}.

\newpage

\begin{figure}[!htb]
\centering
\includegraphics[scale=0.5]{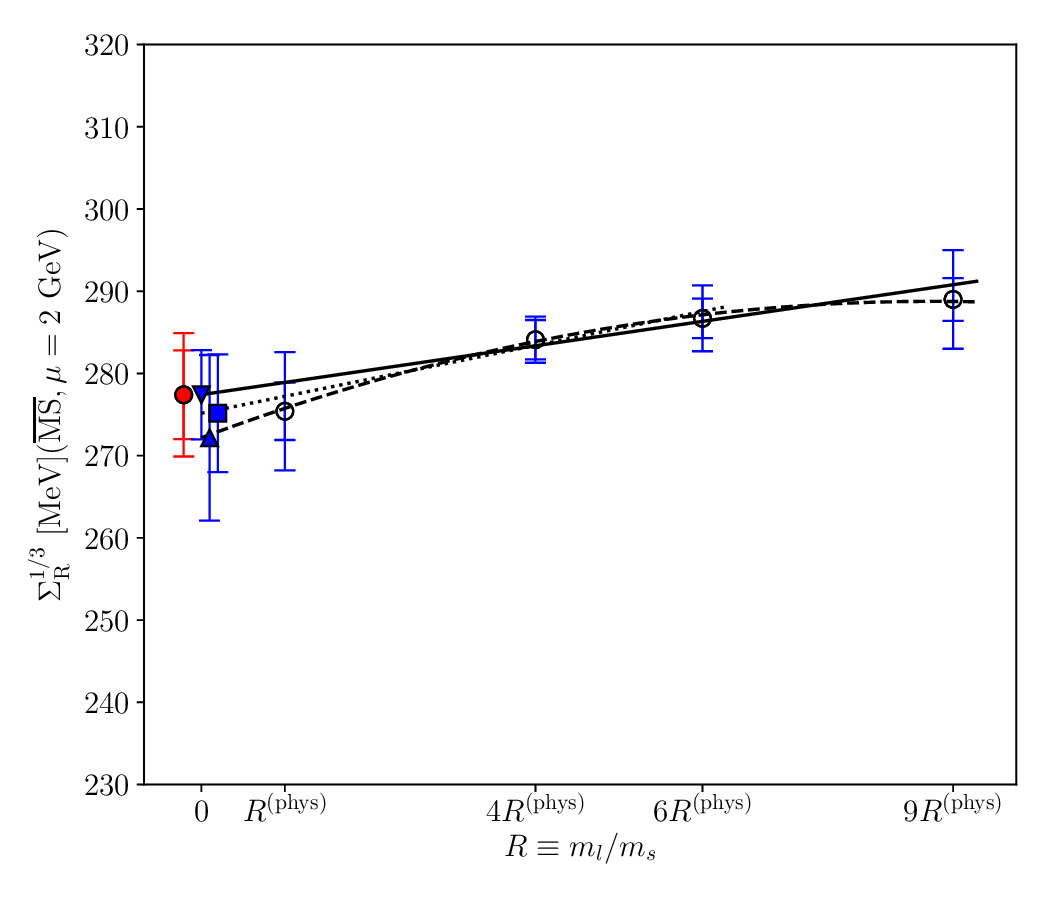}
\caption{Chiral limit extrapolation of $\Sigma_{\mathrm{R}}^{1/3}$ at fixed $m_s$ as a function of $R=m_l/m_s$. For each value of $R$, $\Sigma_\R^{1/3}(R)$ is reported with a double error bar, referring to the statistical and to the sum of statistical and systematic uncertainties. The dotted line represents the result of the best fit performed according to Eq.~\eqref{eq:fit_function_chiral_limit_linear_in_R} and including only the three smallest values of $R$. The straight line is the result from the same fit function, but including also $R=9R^{\phys}$. The dashed line is the best fit result obtained by using a parabolic fit in $R$ in the whole range. The squar point at $R=0$ is the final result from the 3-point fit according to Eq.~\eqref{eq:fit_function_chiral_limit_linear_in_R}, the down-ward triangular point is the one from a linear 4-point fit and the up-ward triangular point is the result obtained from a parabolic 4-point fit. Finally, the round point represents the final value for the chiral condensate extracted from the mode number with a double error bar.}
\label{fig:cc_mode_number_chiral_limit}
\end{figure}

We are now ready to extrapolate our continuum results towards the chiral limit, according to the following fit function~\cite{Giusti:2008vb}:
\beq\label{eq:fit_function_chiral_limit_linear_in_R}
\Sigma_\R^{1/3}(R)=\Sigma_\R^{1/3} + c_2\, R + o(R),
\eeq
where eventually $\Sigma_\R$ represents our final result for the condensate. When performing such extrapolation, we took into account the statistical and systematic errors on the fitted points in the most conservative way, i.e., by considering for each point a single error bar given by the plain sum of the two errors (i.e., assuming they are 100\% correlated). The chiral extrapolation is shown in Fig.~\ref{fig:cc_mode_number_chiral_limit}. As it can be appreciated, our data are perfectly described by a linear function in $R=m_l/m_s$ for all explored values of the light quark mass, as expected from chiral perturbation theory. By fitting all available points, we find $\Sigma_\R^{1/3} = 277.4(5.4)$~MeV. A linear fit to the points corresponding to the three lightest masses gives the perfectly compatible result $\Sigma_\R^{1/3} = 275.2(7.2)$~MeV. Actually, also a parabolic fit in $R$ well describes our data, as it yields the perfectly compatible extrapolation $\Sigma_\R^{1/3} = 272.2(10.1)$~MeV, and a coefficient for $R^2$ which is compatible with zero within errors. In the end, from the mode number we quote the following final result:
\beq
\Sigma_\R^{1/3} = 277.4(5.4)_{\mathrm{stat}}(2.1)_{\mathrm{sys}}~\mathrm{MeV} \qquad \text{ (mode number)},
\eeq
where we took the result of the 4-point linear fit in $R$ for the central value and the statistical error, while the systematic error is determined again according to Eq.~\eqref{eq:systematic_error} from the difference between the chiral limits yielded by the linear 4-point and the quadratic 4-point fits.

Before commenting this number further, we will first proceed with the computation of the chiral condensate with different methods involving different observables, in order to check the consistency of our approach.

\FloatBarrier

\FloatBarrier

\subsection{Chiral condensate from the pion mass}

In this Section we will extract the chiral condensate from the quark mass dependence of the pion mass. From pion correlators, it is possible to extract both the pion mass $M_\pi$ and the pion decay constant $F_\pi$. We will need both quantities to extract the chiral condensate, as, cf.~Eq.~\eqref{eq:GMOR}:
\beq
M_\pi^2 = 2\frac{\Sigma}{F_\pi^2} m_l = 2 \left(\frac{\Sigma m_s}{F_\pi^2}\right) R.
\eeq

\begin{figure}[!b]
\centering
\includegraphics[scale=0.4]{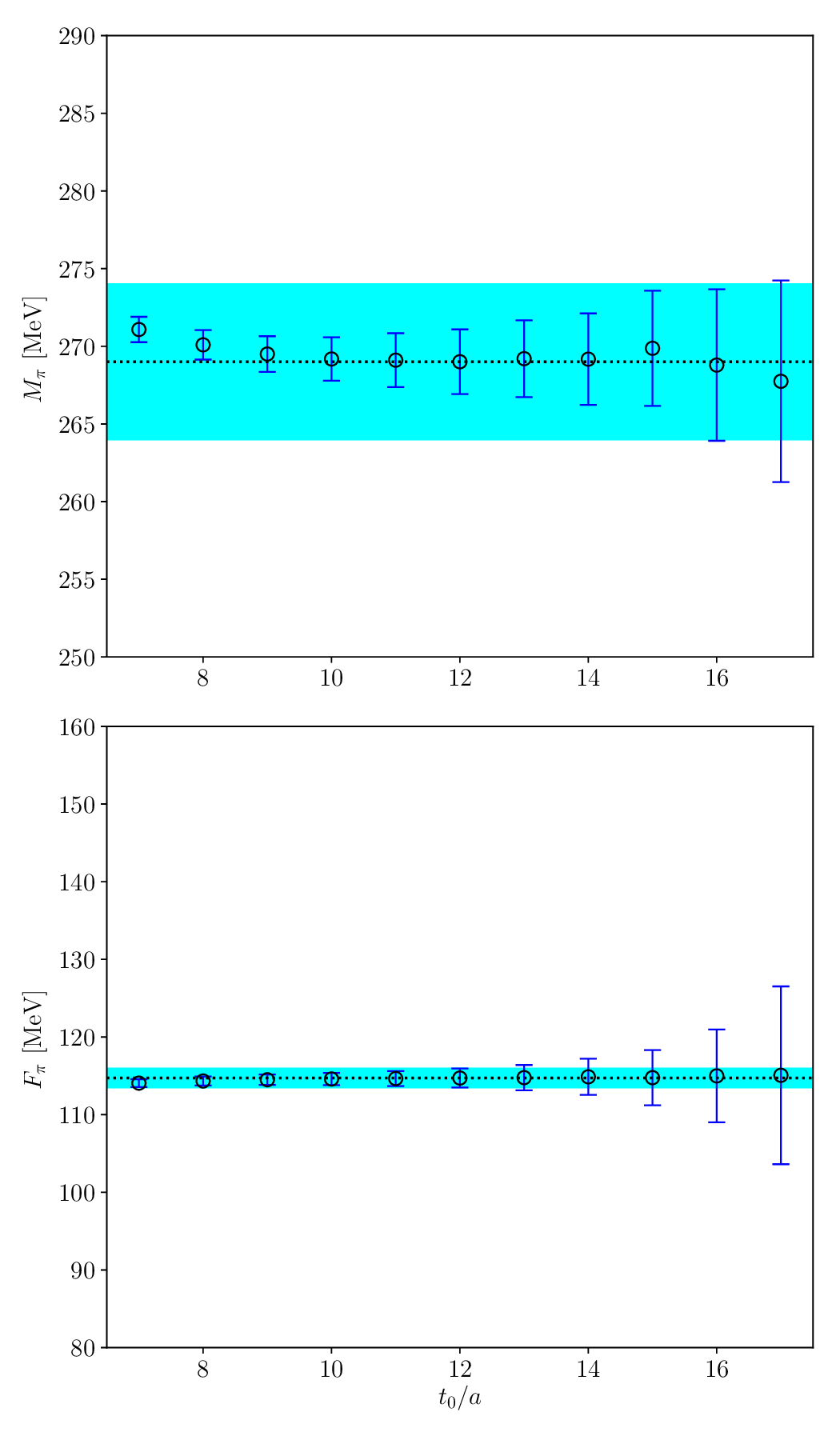}
\caption{Pion mass $M_\pi$ and pion decay constant $F_\pi$ as a function of the lower bound of the fit range $t_0$ (expressed in lattice units) for the ensemble with $N_s=40,\ m_l=4m_l^{\phys}$. Shaded areas represent our final result for both quantities.}
\label{fig:plateau_pion}
\end{figure}

First of all, let us show an example of computation of $M_\pi$ and $F_\pi$. We fitted our pion correlator $C_\pi(t)$ in the range $t/a \in [t_0/a;N_s-t_0/a]$ to the functional form reported in Eq.~\eqref{eq:fit_exp_pion_tcorr} for several values of $t_0$, looking for a pleateau in both quantities, in order to provide a robust estimation of both. The pion decay constant was obtained from Eq.~\eqref{eq:f_pi_from_pion_fit} and using the value of $m_s^{(\R)}$ in Eq.~\eqref{eq_ms_phys}. An example of computation of the pion mass and decay constant is shown in Fig.~\ref{fig:plateau_pion}.

\begin{figure}[!htb]
\centering
\includegraphics[scale=0.4]{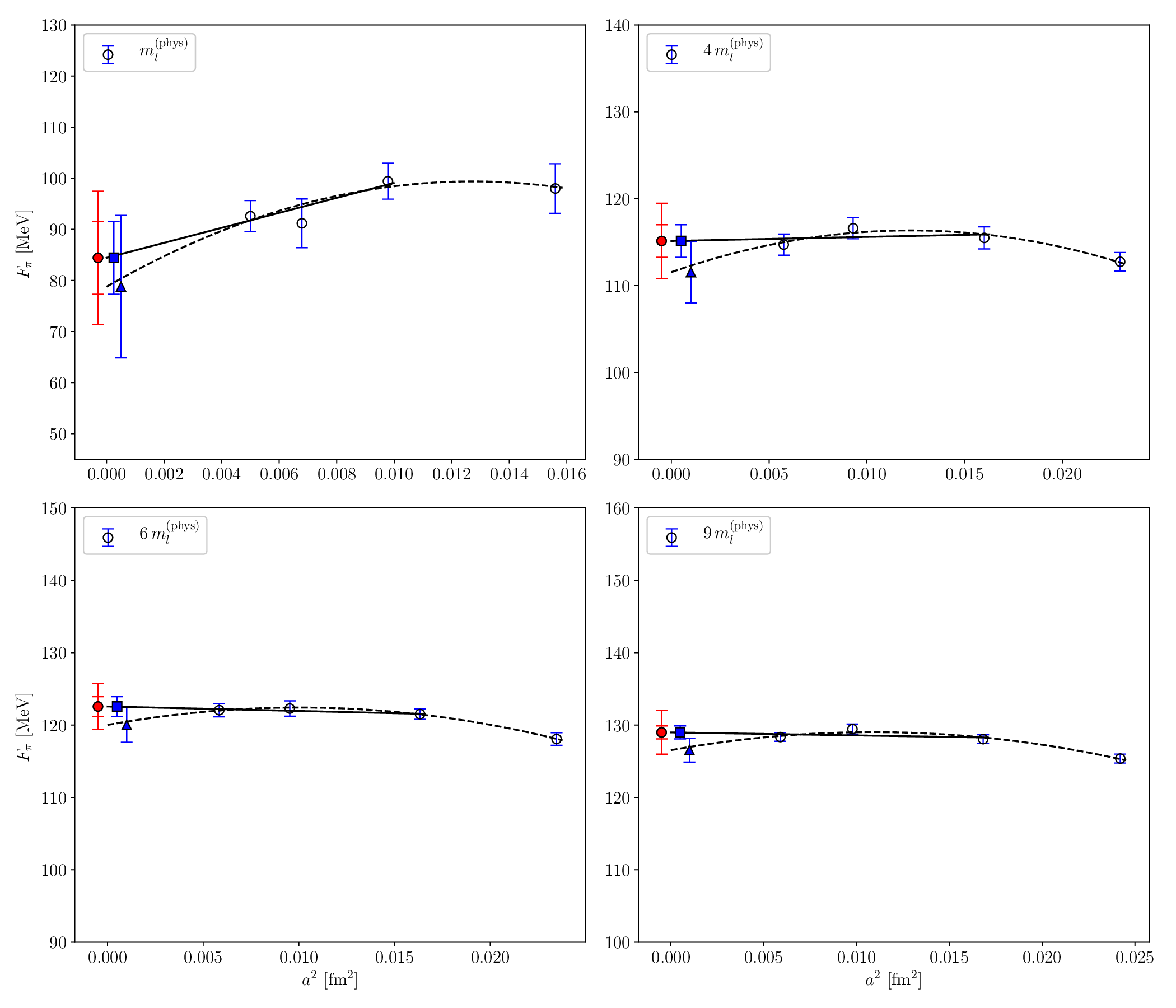}
\caption{Continuum limit extrapolation of $F_\pi(a,R)$ for all the lines of constant physics considered here. Straight and dashed lines represent, respectively, continuum extrapolations assuming $O(a^2)$ corrections to the points corresponding to the three finest lattice spacings and the ones assuming $O(a^4)$ corrections in the whole range. In $a=0$, the squared filled point is the result from the $a^2$-linear fit including the three finest lattice spacings, the triangular one is the $a^2$-parabolic fit result. Finally, the red circled point is the final determination of $F_\pi(R)$, with a double error bar referring to the statistical and to the sum of statistical and systematic uncertainties.}
\label{fig:cont_limit_fpi}
\end{figure}

Let us start our discussion from the pion decay constant. In Fig.~\ref{fig:cont_limit_fpi} we show the extrapolation towards the continuum limit of $F_\pi(a,R)$ at fixed value of $R$ assuming:
\beq
F_\pi(a,R) = F_\pi(R) + b_1(R)\,a^2 + o(a^2).
\eeq
As it can be appreciated, our data can be reliably fitted assuming only $O(a^2)$ corrections when excluding the coarsest lattice spacing, and a parabolic fit in $a^2$ yields perfectly agreeing results when including also the coarsest lattice spacing. Thus, in all cases we took the result of the linear fit for the central values and statistical errors of our continuum extrapolations. A systematic error was also assigned to our continuum limits according to the procedure described in the previous Section. All our continuum determinations for $F_\pi$ are reported in Tab.~\ref{tab:fpi_res}.

\begin{table}[!t]
\begin{center}
\begin{tabular}{ |c|c|c|c|}
\hline
$R\equiv m_l / m_s$ & $m_l / m_l^{\phys}$    & $a$~[fm] & $F_\pi$~[MeV] \\
\hline
\hline
\multirow{5}{*}{\makecell{1/28.15\\$\simeq$ 0.0355}} & \multirow{5}{*}{1} & 0.1249 & 98.0(4.8)\\
&& 0.0989 & 99.4(3.5)\\
&& 0.0824 & 91.2(4.8)\\
&& 0.0707 & 92.6(3.1)\\
&& 0      & $84.4(7.1)_{\mathrm{stat}}(1.8)_{\mathrm{sys}}$\\

\hline	
\hline		
\multirow{5}{*}{\makecell{4/28.15\\$\simeq$ 0.1421}} & \multirow{5}{*}{4} & 0.1515 & 112.7(1.1)\\
&& 0.1265 & 115.5(1.3)\\
&& 0.0964 & 116.6(1.2)\\
&& 0.0758 & 114.7(1.2)\\
&& 0      & $115.1(1.9)_{\mathrm{stat}}(2.5)_{\mathrm{sys}}$\\
\hline
\hline		
\multirow{5}{*}{\makecell{6/28.15\\$\simeq$ 0.2131}} & \multirow{5}{*}{6} & 0.1532 & 118.1(0.9)\\
&& 0.1278 & 121.5(0.7)\\
&& 0.0976 & 122.3(1.1)\\
&& 0.0764 & 122.1(0.9)\\
&& 0      & $122.6(1.4)_{\mathrm{stat}}(1.8)_{\mathrm{sys}}$\\
\hline
\hline
\multirow{5}{*}{\makecell{9/28.15\\$\simeq$ 0.3197}} & \multirow{5}{*}{9} & 0.1556 & 125.4(0.6)\\
&& 0.1297 & 128.1(0.6)\\
&& 0.0989 & 129.4(0.7)\\
&& 0.0768 & 128.3(0.6)\\
&& 0      & $129.0(0.9)_{\mathrm{stat}}(2.1)_{\mathrm{sys}}$\\
\hline
\end{tabular}
\end{center}
\caption{Values of the pion decay constant as a function of the lattice spacing $a$ and of the light quark mass $m_l$. The values of $F_\pi$ at $a=0$ correspond to the extrapolations towards the continuum limit, with statistical and systematic uncertainties.}
\label{tab:fpi_res}
\end{table}

We now proceed to extrapolate our continuum results for $F_\pi$ towards the chiral limit, using the following fit function modeled on ChPT~\cite{Borsanyi:2012zv}:
\beq
F_\pi(R) = F_\pi + b_2\,R + o(R).
\eeq
A linear fit in $R$ to the determinations obtained for the 3 lightest quark masses gives $F_\pi=84.8(8.8)~\mathrm{MeV}$. A parabolic fit in $R$ performed in the whole range gives instead the perfectly compatible result $F_\pi=74.7(11.7)~\mathrm{MeV}$. Thus, as our final result, we take the chiral limit obtained from the linear fit extrapolation, and assign it a systematic uncertainty computed again according to~\eqref{eq:systematic_error}:
\beq\label{eq:pion_decay_constant_final}
F_\pi = 84.8(8.8)_{\mathrm{stat}}(6.1)_{\mathrm{sys}}~\mathrm{MeV}.
\eeq

\begin{figure}[!htb]
\centering
\includegraphics[scale=0.5]{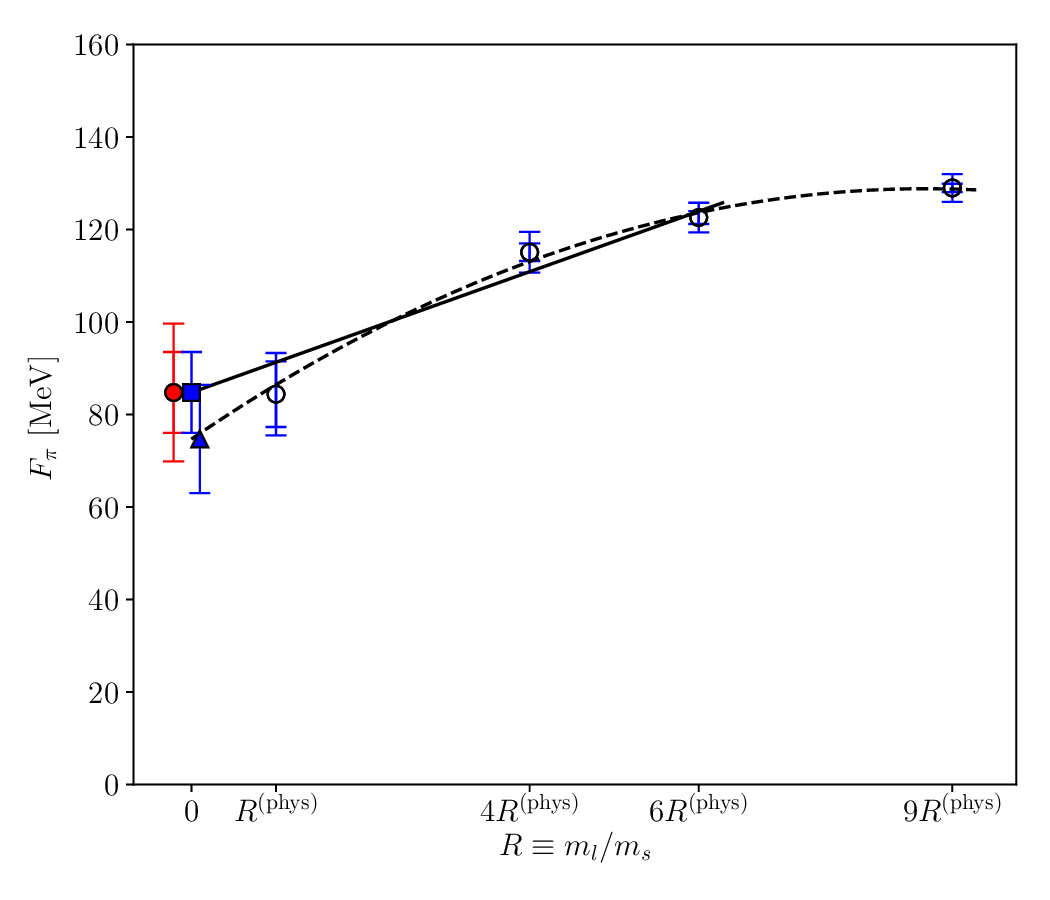}
\caption{Extrapolation towards the chiral limit of $F_\pi(R)$ as a function of $R$. Each value of $F_\pi(R)$ has a double error bar, referring to the statistical and to the sum of the statistical and systematic uncertainties. The straight line represents a linear best fit in $R$ excluding the point for $R=9R^{\phys}$. The dashed line refers to the result from a quadratic fit in $R$ including all available points. The square point at $R=0$ is the determination of $F_\pi$ from the linear fit, the triangular point the one from the parabolic fit. Finally, the round point with the double error bar is the final estimation for $F_\pi$.}
\label{fig:fpi_chiral_lim}
\end{figure}

\begin{figure}[!htb]
\centering
\includegraphics[scale=0.54]{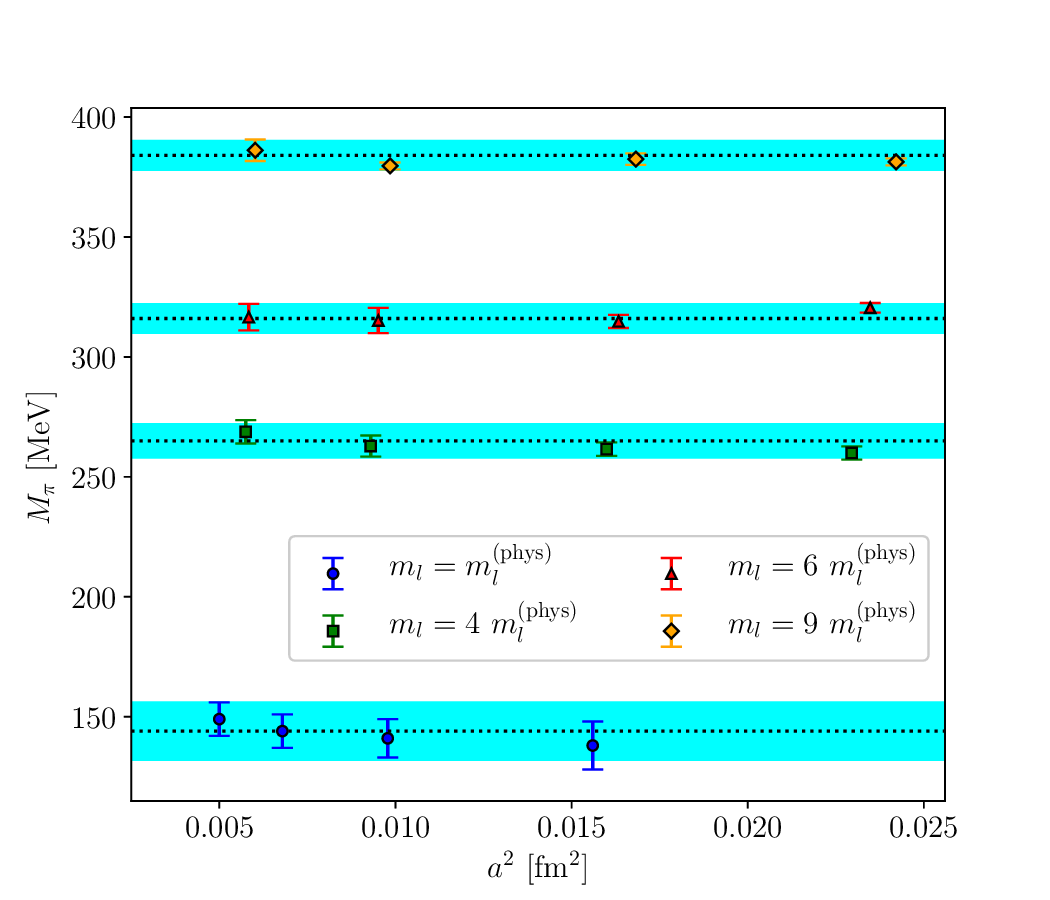}
\caption{Obtained results for the pion mass $M_\pi$ as a function of the lattice spacing for each LCP. The shaded areas represent our final estimations of the pion masses for each LCP.}
\label{fig:pion_mass}
\end{figure}

We now move to our results for $M_\pi$. Pion masses as a function of the lattice spacing are shown in Fig.~\ref{fig:pion_mass}. As expected, they are fairly constant as a function of $a$ for each LCP. The shaded areas in the figure represent our final results for $M_\pi$ for each value of $R$. All our determinations of $M_\pi$ are collected in Tab.~\ref{tab:pion_mass_values}.

\FloatBarrier

\begin{table}[!htb]
\begin{center}
\begin{tabular}{ |c|c|c|c|c|}
\hline
$R\equiv m_l / m_s$ & $m_l / m_l^{\phys}$    & $a$~[fm] & $M_\pi(a)$~[MeV] &  \makecell{Final res.\\$M_\pi$~[MeV]}\\
\hline
\hline
\multirow{4}{*}{\makecell{1/28.15\\$\simeq$ 0.0355}} & \multirow{4}{*}{1} & 0.1249 & 138(10) & \multirow{4}{*}{144(12)}\\
&& 0.0989 & 141(8) &\\
&& 0.0824 & 144(7) &\\
&& 0.0707 & 149(7) &\\
\hline	
\hline		
\multirow{4}{*}{\makecell{4/28.15\\$\simeq$ 0.1421}} & \multirow{4}{*}{4} & 0.1515 & 260(3) & \multirow{4}{*}{265(6)}\\
&& 0.1265 & 262(3) &\\
&& 0.0964 & 263(4) &\\
&& 0.0758 & 269(5) &\\
\hline
\hline		
\multirow{4}{*}{\makecell{6/28.15\\$\simeq$ 0.2131}} & \multirow{4}{*}{6} & 0.1532 & 320(2) & \multirow{4}{*}{316(6)}\\
&& 0.1278 & 315(3) &\\
&& 0.0976 & 315(5) &\\
&& 0.0764 & 317(6) &\\
\hline
\hline
\multirow{4}{*}{\makecell{9/28.15\\$\simeq$ 0.3197}} & \multirow{4}{*}{9} & 0.1556 & 381(1) & \multirow{4}{*}{384(7)}\\
&& 0.1297 & 382(2) & \\
&& 0.0989 & 380(1) & \\
&& 0.0768 & 386(4) & \\
\hline
\end{tabular}
\end{center}
\caption{Values of the pion mass $M_\pi$ for each value of $m_l$ and for each lattice spacing $a$.}
\label{tab:pion_mass_values}
\end{table}

\begin{figure}[!htb]
\centering
\includegraphics[scale=0.58]{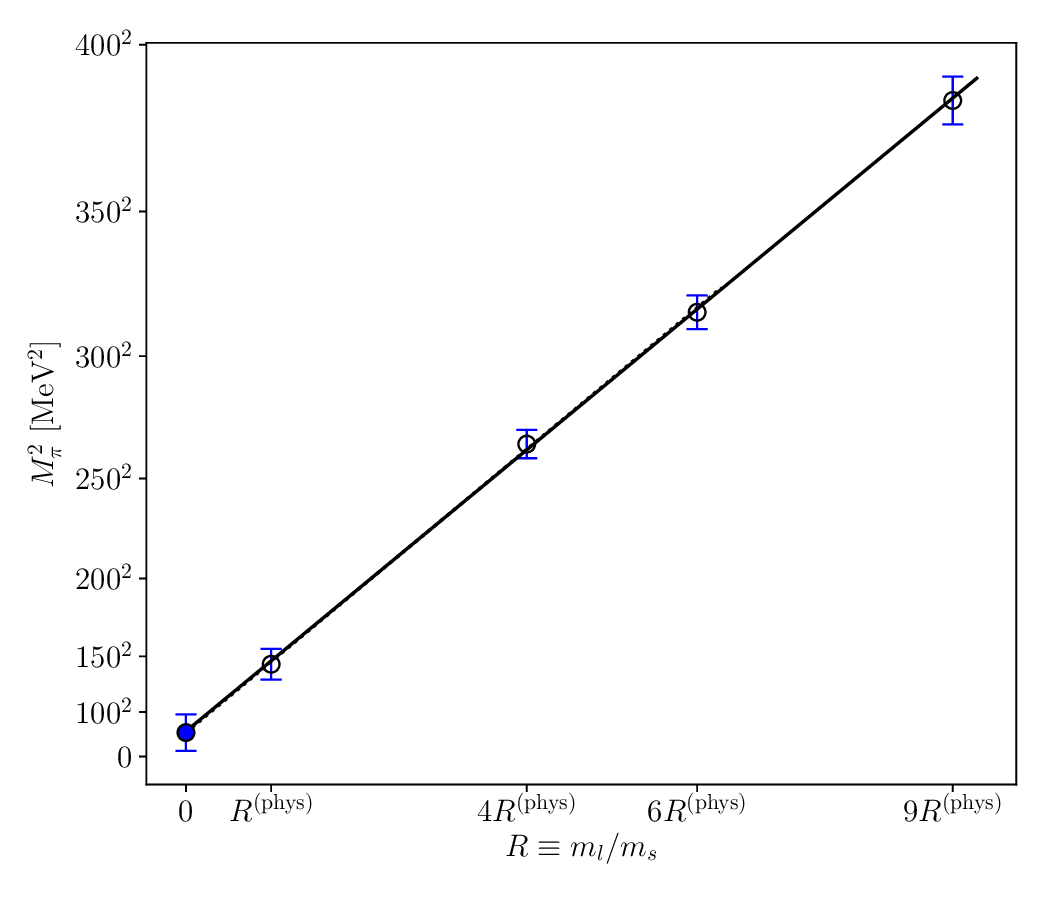}
\caption{Extrapolation towards the chiral limit of $M_\pi^2$ at fixed $m_s$ as a function of $R=m_l/m_s$. The straight line is the result of a linear fit where the chiral limit  of $M_\pi^2$ is left as a free parameter. The dotted line, nearly indistinguishable from the solid one, is the best fit with respect to the same function but excluding the heavier pion mass. The filled point at $R=0$ is the chiral limit result including the whole range. In both cases, we find such fit parameter to be compatible with zero within errors.}
\label{fig:mpi_vs_R}
\end{figure}

As it can be appreciated from Fig.~\ref{fig:mpi_vs_R}, our results for $M_\pi^2$ can be nicely fitted with a linear function in $R$. First, we perform a linear fit where the chiral limit of the pion mass is left as a free parameter. We find, using the number in Eq.~\eqref{eq:pion_decay_constant_final} for the pion decay constant, $\Sigma_\R^{1/3} = 258.7(30.6)$~MeV if all available points are included in the fit, and $\Sigma_\R^{1/3} = 259.3(30.9)$~MeV if the heaviest pion is excluded. In both cases we find a vanishing chiral limit for $M_\pi$ within errors. As a matter of fact, if we repeat this fit fixing the chiral limit of $M_\pi$ to zero, we find $\Sigma_\R^{1/3} = 264.0(31.0)$~MeV and $\Sigma_\R^{1/3} = 265.4(31.2)$~MeV if the heaviest pion is included/excluded from the fit respectively. In the end, from the pion mass we quote the final result:
\beq
\Sigma_\R^{1/3} = 258.7(30.6)_{\mathrm{stat}}(0.01)_{\mathrm{sys}}~\mathrm{MeV} \qquad \text{ (pion mass)},
\eeq
where the systematic uncertainty, again computed according to Eq.~\eqref{eq:systematic_error}, turns out to be very small, being the differences among the chiral limits obtained from the 3- and 4-point fits extremely small compared to their statistical errors.

\FloatBarrier

\subsection{Chiral condensate from the topological susceptibility}

In this Section we will address the computation of the chiral condensate from the quark mass dependence of the topological susceptibility, according to the ChPT prediction in Eq.~\eqref{eq:ChPT_topsusc}:
\beq
\chi = \frac{1}{2} \Sigma m_l = \frac{1}{2} (\Sigma m_s) R.
\eeq

To keep the discussion more compact, here we will just show the computation of the continuum limit of $\chi$ for one line of constant physics, namely the one corresponding to $m_l = 4 m_l^{\phys}$. Concerning the computation of $\chi$ at the physical point, it has been already extensively discussed on the dedicated paper~\cite{Athenodorou:2022aay}, thus, here we just take the result reported there. Finally, the results for the other two lines of constant physics with heavier-than-physical pions can be found in App.~\ref{app:additional_plots}.

First of all, we extrapolate our results towards the continuum limit using the fermionic discretization discussed in Sec.~\ref{sec:topsusc}, and assuming the following fit function:
\beq
\chi_\SP^{1/4}(a,M/m_s) &=& \chi^{1/4} + c_\SP(M/m_s) \, a^2 + o(a^2).
\eeq

\begin{figure}[!htb]
\centering
\includegraphics[scale=0.42]{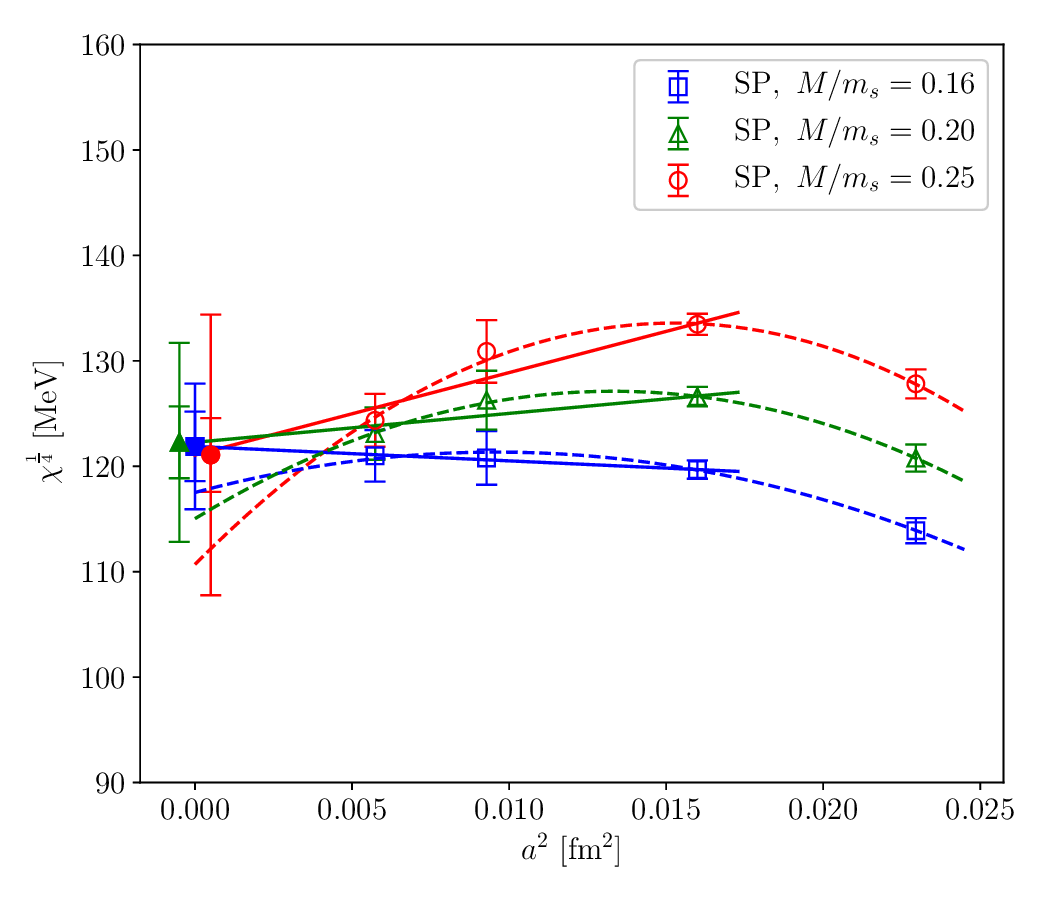}
\includegraphics[scale=0.42]{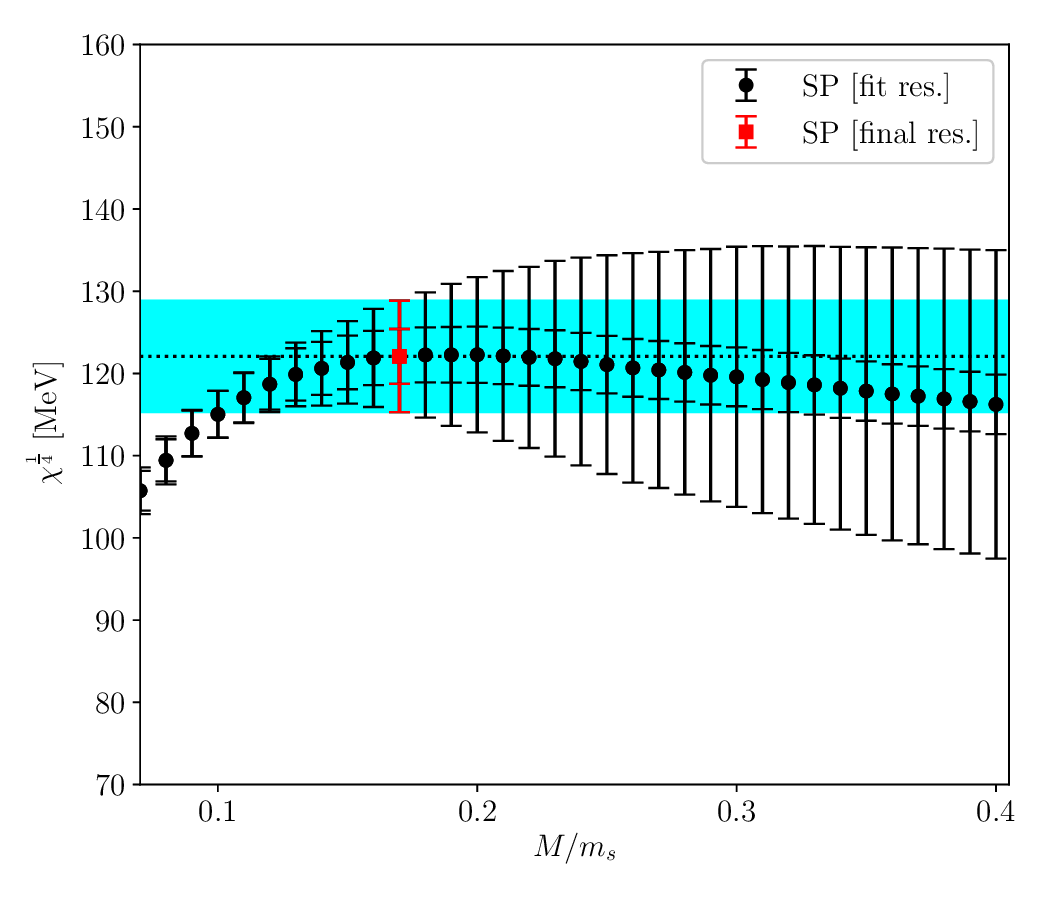}
\caption{Left panel: comparison of continuum limit extrapolations of $\chi_\SP^{1/4}(M/m_s)$ for $m_l=4m_l^{(\mathrm{phys})}$ and for a few values of $M/m_s$. Straight lines refer to the continuum extrapolations obtained by fitting determinations obtained for the 3 finest lattice spacings with a linear function in $a^2$. Dashed lines refer to the best fit results with a $a^2$-parabolic fit function. Full points in $a=0$ represent the corresponding continuum extrapolations. Right panel: continuum limits of $\chi_{\SP}^{1/4}$ obtained from spectral
projectors at $m_l=4m_l^{(\mathrm{phys})}$ for several values of $M/m_s$. The square point and the shaded area represent our final result for the continuum limit of $\chi_\SP$. The double error bar convention is the same of the previous plots.}
\label{fig:comparison_spectral_topsusc_4ml}
\end{figure}

In Fig.~\ref{fig:comparison_spectral_topsusc_4ml} on the left, we show a few examples of continuum extrapolations for some values of $M/m_s$. As it can be observed, our spectral definition suffers for mild lattice artifacts if $M/m_s$ is taken sufficiently small, and our determinations for the 3 finest lattice spacings can be reliably fitted with a linear function in $a^2$. We also tried to perform parabolic fits in $a^2$ including also the determinations at the coarsest lattice spacing, obtaining agreeing results within errors.  Thus, for each value of $M/m_s$, we took the linear extrapolations obtained with 3 fitted points as our estimations of the continuum limit of $\chi^{1/4}_\SP(M/m_s)$. systematic errors were assigned to the continuum extrapolations by comparing the linear and the parabolic fits with the method introduced in Sec.~\ref{sec:sigma_from_mode_number}.

In Fig.~\ref{fig:comparison_spectral_topsusc_4ml} on the right we show instead how the continuum extrapolations of $\chi_\SP^{1/4}(M/m_s)$ behave as a function of $M/m_s$, with each point having a double error bar that refers to both the statistical and the sum of the statistical and systematic uncertainties. On general theoretical grounds, we expect the continuum limit of $\chi_\SP(M/m_s)$ to be independent of $M/m_s$. As it can be observed, continuum extrapolations of $\chi_\SP^{1/4}(M/m_s)$ as a function of $M/m_s$ are all compatible among each other, as expected, with a systematic uncertainty growing as the threshold mass is increased. As already discussed in detail in Ref.~\cite{Athenodorou:2022aay}, this is due to the fact that, as $M/m_s$ is increased, we are including more and more non-chiral modes in our spectral sums. This makes lattice artifacts grow larger, eventually reaching the same magnitude of those affecting the gluonic definition. As a consequence, the continuum extrapolation is affected by larger systematic effects for larger values of $M/m_s$. In the end, as our final value for the continuum limit of the susceptibility, we took a point within the plateau that is clearly visible in the right panel of Fig.~\ref{fig:comparison_spectral_topsusc_4ml} for $M/m_s\gtrsim 0.14$. Our final continuum results are reported in Tab.~\ref{tab:toposusc_values}.

\begin{table}[!htb]
\begin{center}
\begin{tabular}{ |c|c|c|c|}
\hline
$R\equiv m_l / m_s$ & $m_l / m_l^{\phys}$ & $\chi^{1/4}$~[MeV] \\
\hline
1/28.15 $\simeq$ 0.0355 & 1 & $80.0(4.0)_{\mathrm{stat}}(8.0)_{\mathrm{sys}}$ \\
\hline
4/28.15 $\simeq$ 0.1421 & 4 & $122.1(3.3)_{\mathrm{stat}}(3.4)_{\mathrm{sys}}$ \\
\hline	
6/28.15 $\simeq$ 0.2131 & 6 & $127.0(4.9)_{\mathrm{stat}}(3.3)_{\mathrm{sys}}$ \\
\hline
9/28.15 $\simeq$ 0.3197 & 9 & $142.5(9.5)_{\mathrm{stat}}(2.7)_{\mathrm{sys}}$\\
\hline
\end{tabular}
\end{center}
\caption{Final continuum limit determinations of the fourth root of the topological susceptibility, obtained with the spectral projectors discretization. The first error is statistical, the second one takes into account all sources of systematic uncertainties (see the text for more details).}
\label{tab:toposusc_values}
\end{table}

\begin{figure}[!htb]
\centering
\includegraphics[scale=0.44]{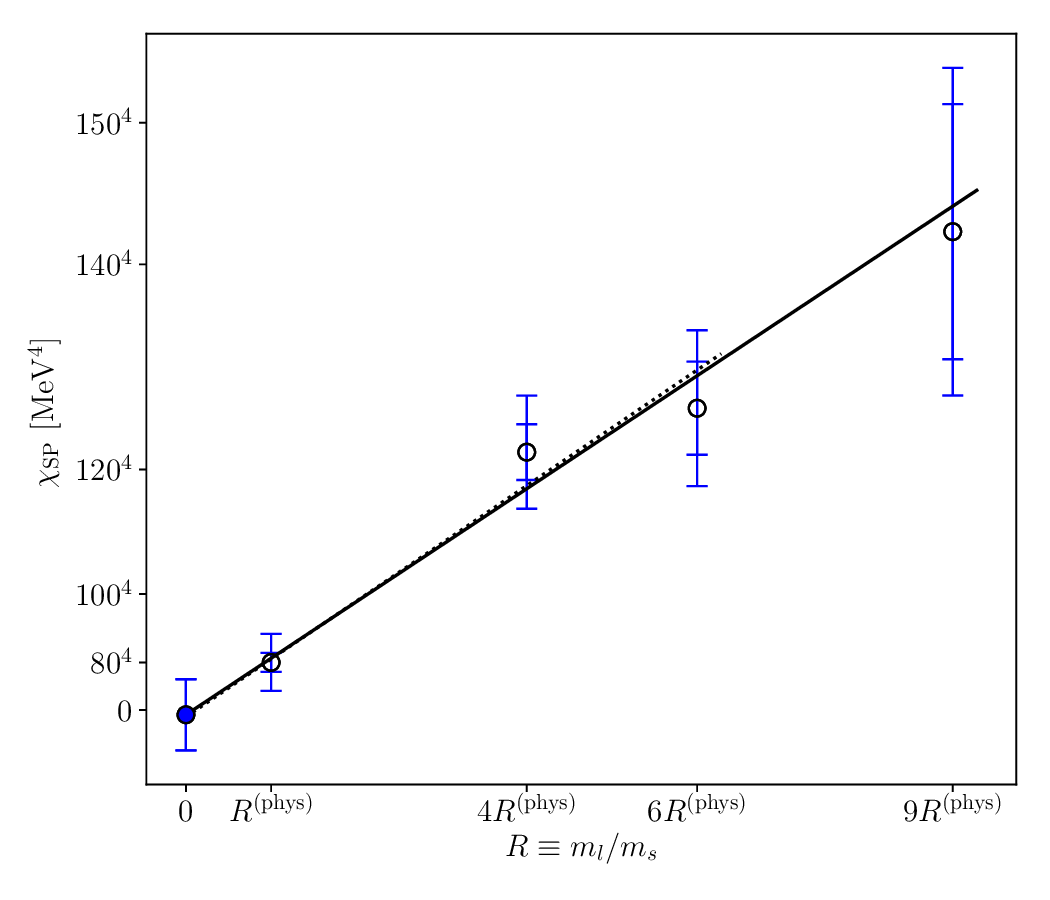}
\caption{Extrapolation towards the chiral limit of the continuum results for the topological susceptibility $\chi$ at fixed $m_s$ as a function of $R=m_l/m_s$. The solid line is the result of a linear fit with the chiral limit left as a free parameter including the whole range. The dotted line refers to the best fit result obtained with the same function but excluding the point at $R=9R^{\phys}$. We find such fit parameter to be compatible with zero within errors. We adopt alwawys the same convention for the double error bar.}
\label{fig:chiral_limit_toposusc}
\end{figure}

As it can be appreciated from Fig.~\ref{fig:chiral_limit_toposusc}, our results for the topological susceptibility can be perfectly described by a linear function in $R$. First, we perform a linear fit where the chiral limit of $\chi$ is left as a free parameter. We find $\Sigma_\R^{1/3} = 309.6(22.1)$~MeV if all available points are included in the fit, and $\Sigma_\R^{1/3} = 312.0(25.2)$~MeV if the point with $R=9R^{\phys}$ is excluded. In both cases we find a vanishing chiral limit for $\chi$ within errors. As a matter of fact, if we repeat this fit fixing the chiral limit of $\chi$ to zero, we find $\Sigma_\R^{1/3} = 308.4(16.8)$~MeV and $\Sigma_\R^{1/3} = 307.4(15.1)$~MeV if the point at $9R^{\phys}$ is included/excluded from the fit respectively. In the end, from the topological susceptibility we quote the final result:
\beq
\Sigma_\R^{1/3} = 309.6(22.1)_{\mathrm{stat}}(0.2)_{\mathrm{sys}}~\mathrm{MeV} \qquad \text{ (topological susceptibility)},
\eeq
where the central value and the statistical error are the ones obtained from the 4-point linear fit, while the very small systematic one comes from the comparison between the results of the 3-point and the 4-point fits, and was computed again using~\eqref{eq:systematic_error}.

\subsection{Discussion of the obtained results and global fit}

So far, we have obtained these three different determinations of the chiral condensate:
\beq
\Sigma_\R^{1/3} &= 277.4(5.4)_{\mathrm{stat}}(2.1)_{\mathrm{sys}}~\mathrm{MeV}  \qquad &\text{ (from $\braket{\nu}$)},\\
\Sigma_\R^{1/3} &= 258.7(30.6)_{\mathrm{stat}}(0.01)_{\mathrm{sys}}~\mathrm{MeV} \qquad &\text{ (from $M_\pi$)},\\
\Sigma_\R^{1/3} &= 309.6(22.1)_{\mathrm{stat}}(0.2)_{\mathrm{sys}}~\mathrm{MeV} \qquad &\text{ (from $\chi$)}.
\eeq
Since they all are in very good agreement among themselves, it is reasonable to perform a global fit of these quantities as a function of the ratio of quark masses $R$ to provide a more stringent check of the consistency of these findings among themselves.

\begin{figure}[!t]
\centering
\includegraphics[scale=0.5]{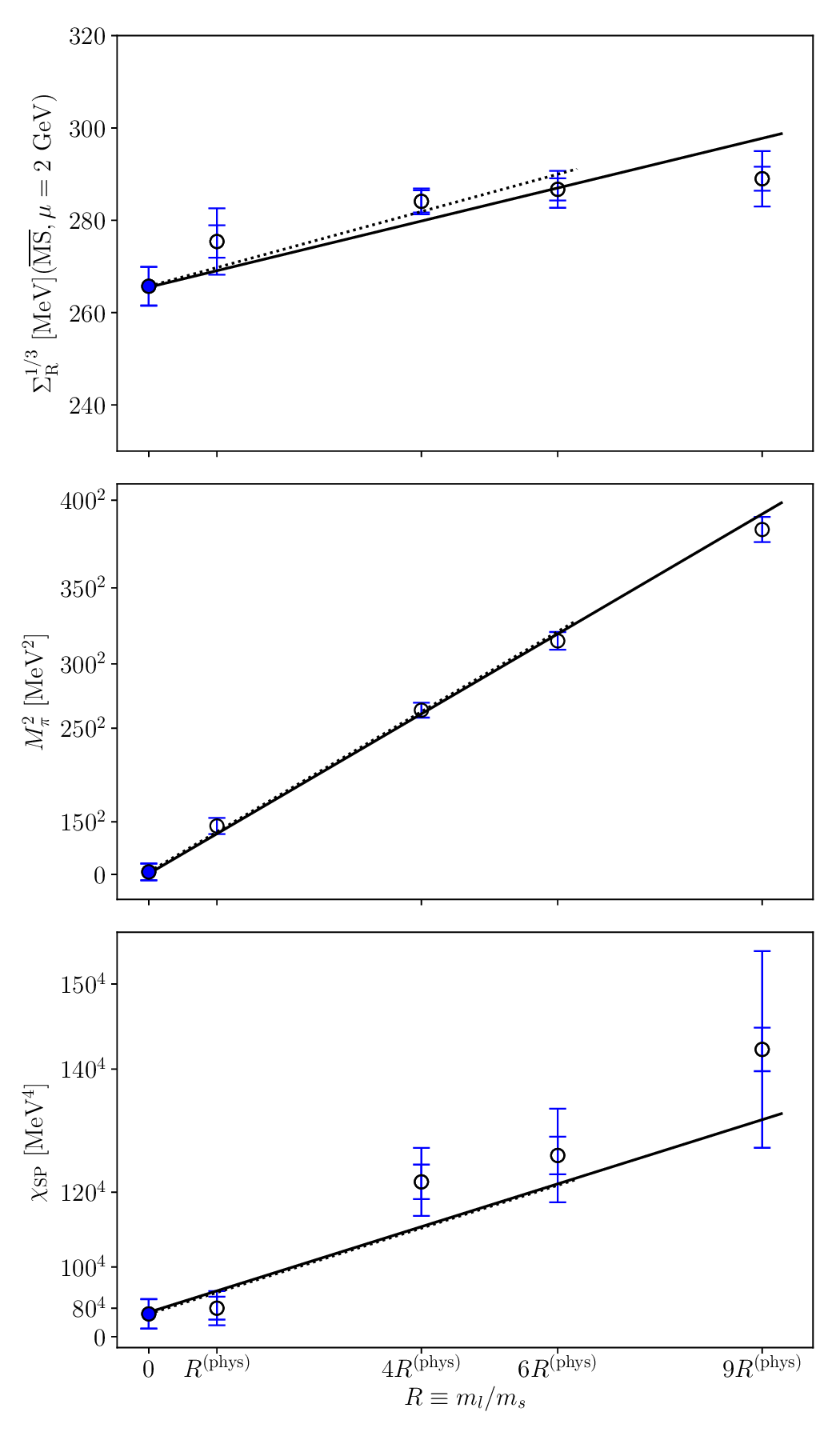}
\caption{Result of the global chiral fit to the continuum limits of the effective condensate obtained from the mode number (top plot), the continuum limits of $M_\pi^2$ (center plot) and the continuum limits of the topological susceptibility, obtained using the spectral projectors definition (bottom plot), imposing that all data sets are described by a single free fit parameter for the chiral condensate, and leaving the chiral limits of $M_\pi$ and $\chi$ as free fit parameters. Straight lines refer to the fit results obtained with all available points, while dotted one to the results obtained with the same fit function but excluding the point $R=9R^{\phys}$ The double error bar convention is used.}
\label{fig:global_fit}
\end{figure}

To this end, we perform a fit of the continuum limits of the effective condensate $\Sigma_\R(R)$ extracted from the mode number, of $M_\pi^2(R)$ extracted from pion correlators and of the continuum limits of the spectral topological susceptibility $\chi(R)$, imposing a single fit parameter for the chiral condensate $\Sigma_\R$, and using our result in Eq.~\eqref{eq:pion_decay_constant_final} for the pion decay constant. Leaving the chiral limit of $M_\pi$ and $\chi$ as free parameters, and including all available determinations in the best fit procedure, we find that our data can be perfectly described by such global fit, just involving one free parameter for the chiral condensate. Such fit is shown in Fig.~\ref{fig:global_fit}. It yields a reduced chi-squared $\tilde{\chi}/\mathrm{dof}=12.16/8$, vanishing chiral limits for $M_\pi$ and $\chi$ within errors, and a value for the chiral condensate $\Sigma_\R^{1/3} = 265.5(3.1)$~MeV. Excluding from the global fit our determinations obtained for the heaviest pion here considered, we find the perfectly agreeing result $\Sigma_\R^{1/3} = 265.7(4.2)$~MeV. Also fixing the chiral limits of the pion mass and of the topological susceptibility to be zero does not change the obtained results, as we find: $\Sigma_\R^{1/3} = 265.9(1.8)$~MeV and $\Sigma_\R^{1/3} = 267.0(2.4)$~MeV if, respectively, quantities obtained for $R=9R^{\phys}$ are included/excluded from the fit. Thus, in the end, we quote the following result for the chiral condensate from the global fit, which we also take as our final determination for $\Sigma_\R$:
\beq\label{eq:final_res}
\Sigma_\R^{1/3} = 265.7(4.2)_{\mathrm{stat}}(0.5)_{\mathrm{sys}}~\mathrm{MeV}, \qquad \text{ (global fit)}.
\eeq
The central value and the associated statistical uncertainty come from the 4-point best fit result, with the chiral limits of $M_\pi$ and $\chi$ left as free parameters. The systematic error, instead, was compued from Eq.~\eqref{eq:systematic_error} from the difference between the chiral extrapolations obtained when fixing $M_\pi(R=0)$ and $\chi(R=0)$ to zero and when leaving them as free parameters, as the variation observed among the 3-point and the 4-point fits was always negligible.

Although the quantities involved in this fitting procedure have been computed on the same set of configurations, and are thus correlated, we expect such correlations not to have a significant impact, since the fitted observables are very different and affected by very different systematic errors, which have all been taken into account very conservatively. In any case, being the determination from the mode number much more precise than the others, we expect it to drive the result of the global fit. By looking at our final result in Eq.~\eqref{eq:final_res} we observe that this is the case, which makes us confident that the error estimate on our final result for the condensate is overall pretty solid.

As a final remark, let us comment that this result is in perfect agreement both with the phenomenological estimation given in the introduction, $\Sigma_{\mathrm{pheno}}^{1/3} = 283(24)$ MeV, and with the world-average reported in the latest FLAG review for the $\SU(2)$ condensate obtained from $N_f=2+1$ QCD: $\Sigma_{\mathrm{FLAG}}^{1/3} = 272(5)$ MeV~\cite{FlavourLatticeAveragingGroupFLAG:2021npn}.

\FloatBarrier

\section{Conclusions}\label{sec:conclu}

In this paper we have addressed the computation of the $\SU(2)$ chiral condensate from a staggered discretization of $2+1$ QCD.

Our calculation is based on 4 lattice spacings and 4 lines of constant physics with different values of the light quark mass, and the strange quark mass kept at the physical point. This allowed us to perform controlled continuum and chiral extrapolations.

Concerning the numerical strategies pursued in this work, our main approach relies on the extraction of the chiral condensate from the mode number using the Giusti--L\"uscher method, based on the Banks--Casher relation. Such technique is implemented and applied to the staggered case for the first time in this paper.

Moreover, we checked carefully that perfectly agreeing results are obtained if the chiral condensate is extrated from the quark mass dependence of the pion mass or the quark mass dependence of the topological susceptibility, which we have computed from spectral projectors on the low-lying modes of the staggered operator. The agreement among such different determinations is further confirmed by a global fit of our data, assuming a single fit parameter for the chiral condensate, which provides an excellent description of our numerical results. Finally, as a by-product of our study, we are also able to provide an estimate of the pion decay constant in the chiral limit $F_\pi$.

In the end, we provide the following final results for the two $\SU(2)$ ChPT LECs:
\beq
\Sigma_\R^{1/3} &=& 265.7(4.2)_{\mathrm{stat}}(0.5)_{\mathrm{sys}}~\mathrm{MeV},\qquad (\overline{\mathrm{MS}}, \, \mu = 2~\mathrm{GeV}),\\
F_\pi &=& 84.7(8.8)_{\mathrm{stat}}(6.1)_{\mathrm{sys}}~\mathrm{MeV}.
\eeq
Our results are in excellent agreement both with the phenomenological estimation that can be obtained from the GMOR relation, $\Sigma_{\mathrm{pheno}}^{1/3} = 283(24)$ MeV, and with the FLAG $2+1$ results: $\Sigma_{\mathrm{FLAG}}^{1/3} = 272(5)$~MeV and $F_\pi\big\vert_{\mathrm{FLAG}}=86.6(6)$~MeV.

\section*{Acknowledgements}
We thank L.~Giusti for useful discussions. The work of C.~Bonanno is supported by the Spanish Research Agency (Agencia Estatal de Investigación) through the grant IFT Centro de Excelencia Severe Ochoa CEX2020- 001007-S and, partially, by grant PID2021-127526NB-I00, both funded by MCIN/AEI/ 10.13039/ 501100011033. C.~Bonanno also acknowledges support from the project H2020-MSCAITN-2018-813942 (EuroPLEx) and the EU Horizon 2020 research and innovation programme, STRONG-2020 project, under grant agreement No 824093. Numerical simulations have been performed on the \texttt{MARCONI} and \texttt{MARCONI100} machines at CINECA, based on the Project IscrB ChQCDSSP and on the agreement between INFN and CINECA (under projects INF22\_npqcd, INF23\_npqcd).

\appendix
\section*{Appendix}
\section{Additional plots}\label{app:additional_plots}

\begin{figure}[!htb]
\centering
\includegraphics[scale=0.25]{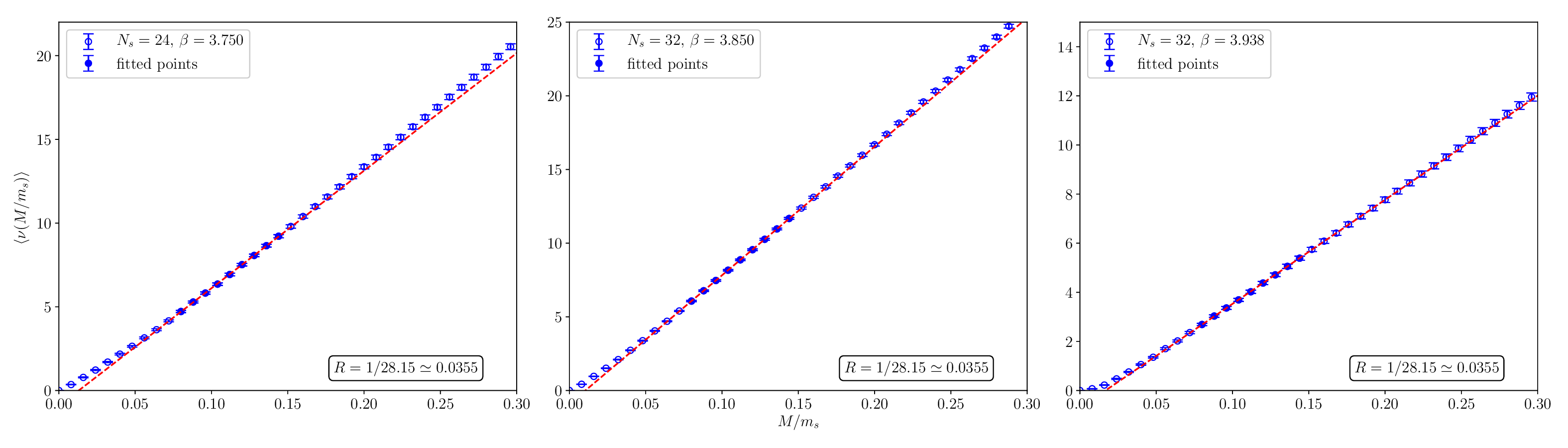}
\includegraphics[scale=0.25]{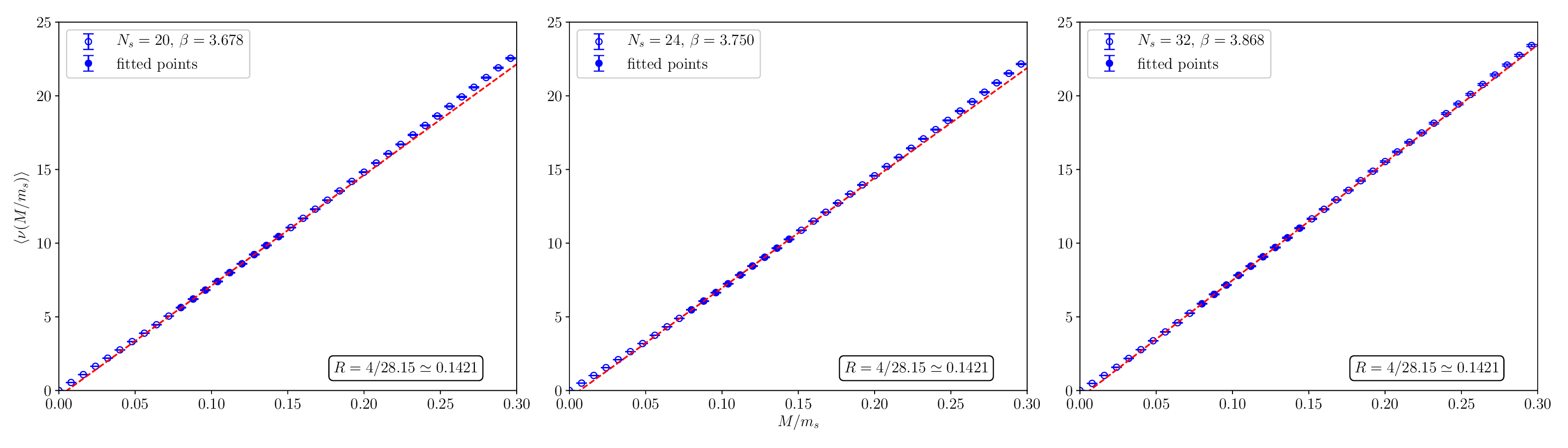}
\includegraphics[scale=0.25]{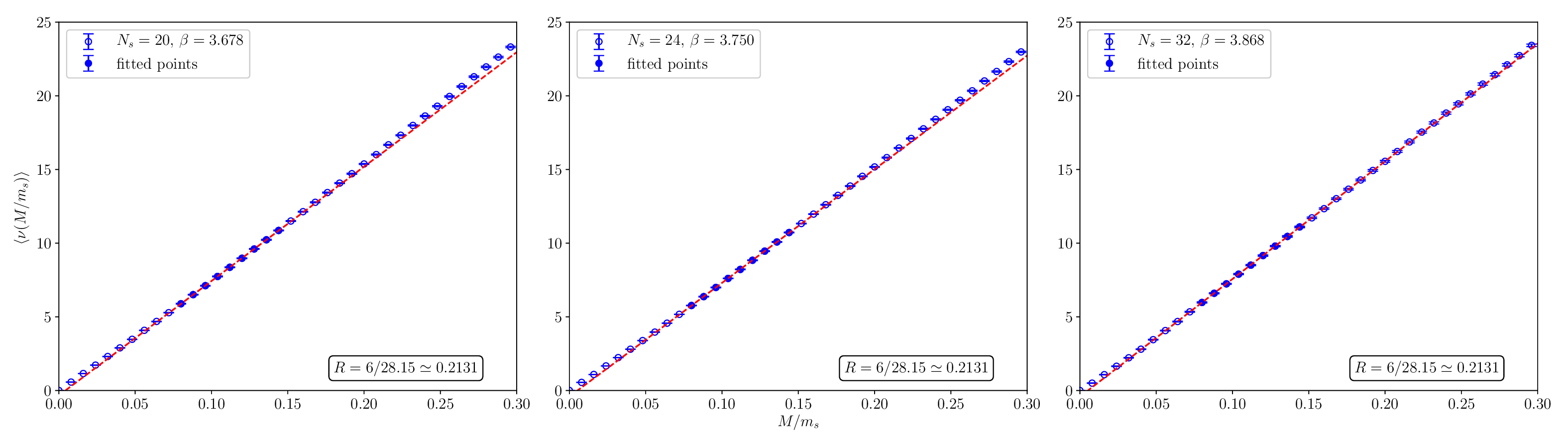}
\includegraphics[scale=0.25]{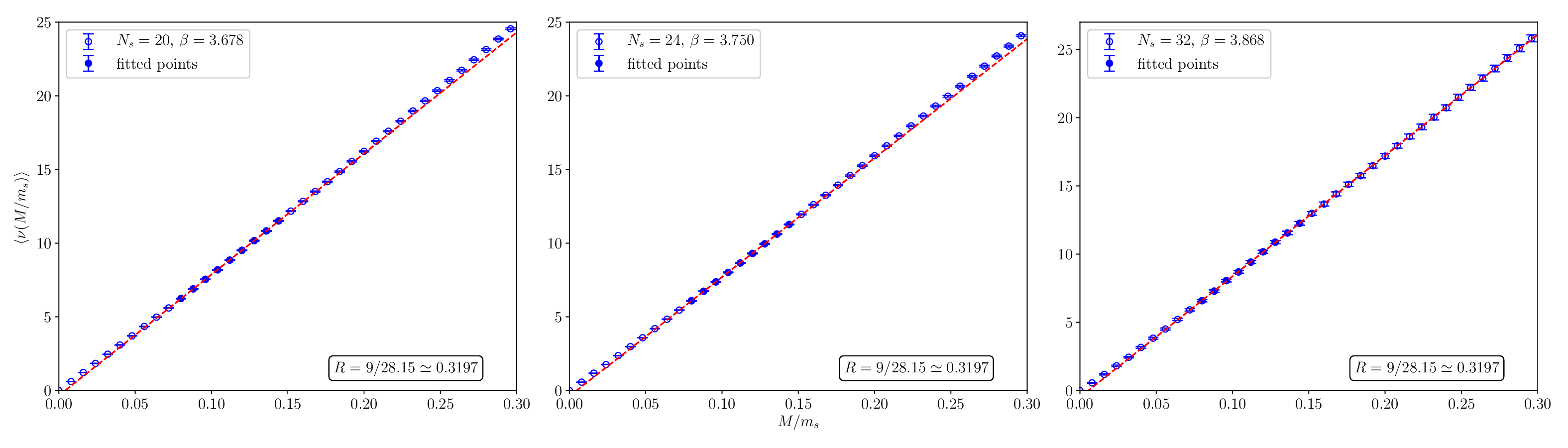}
\caption{Linear best fits of the physical mode number $\braket{\nu} = \braket{\nu_\stag}/4$ as a function of $M/m_s$ for the 3 coarser lattice spacings at all values of the pion mass. Filled points in the range $M/m_s\in[0.075,0.15]$ are those included in the best fits, depicted as dashed lines.}
\label{fig:fit_mode_number_appendix}
\end{figure}

\begin{figure}[!htb]
\centering
\includegraphics[scale=0.42]{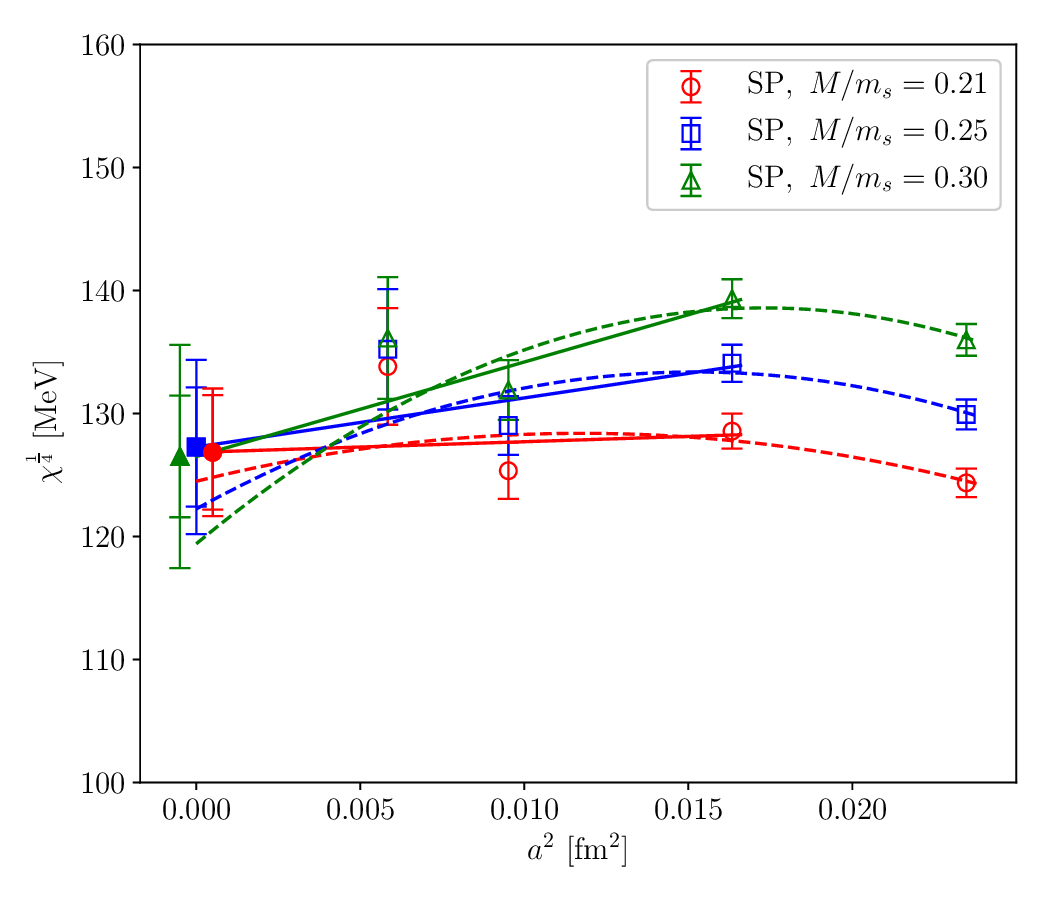}
\includegraphics[scale=0.42]{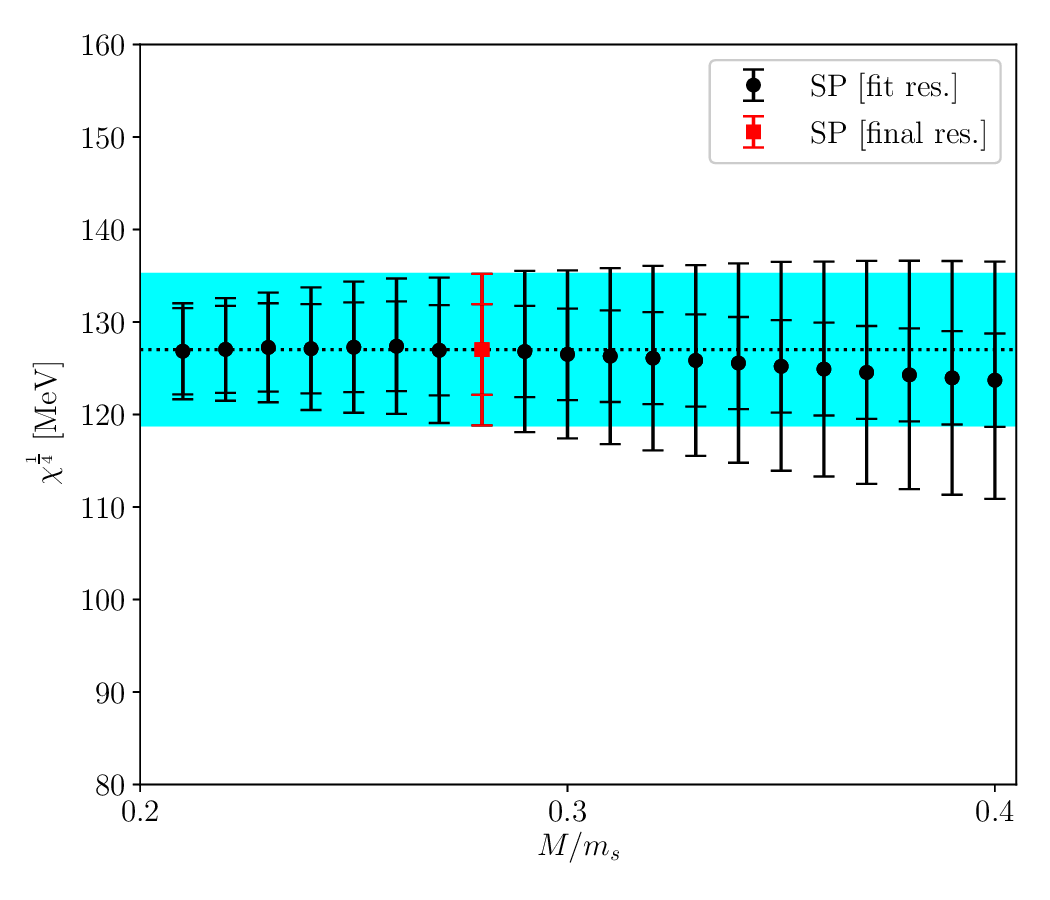}
\caption{Left panel: comparison of continuum limit extrapolations of $\chi_\SP^{1/4}(M/m_s)$ for $m_l=6m_l^{(\mathrm{phys})}$ and for a few values of $M/m_s$. Straight lines refer to the continuum extrapolations obtained by fitting determinations obtained for the 3 finest lattice spacings with a linear function in $a^2$. Dashed lines refer to the best fit result with a $a^2$-parabolic fit function. Full points in $a=0$ represent the corresponding continuum extrapolations. Right panel: continuum limits of $\chi_{\SP}^{1/4}$ obtained from spectral
projectors at $m_l=6m_l^{(\mathrm{phys})}$ for several values of $M/m_s$. The square point and the shaded area represent our final result for the continuum limit of $\chi_\SP$. The double error bar convention is the same of the previous plots.}
\label{fig:comparison_spectral_topsusc_6ml}
\end{figure}

\begin{figure}[!htb]
\centering
\includegraphics[scale=0.42]{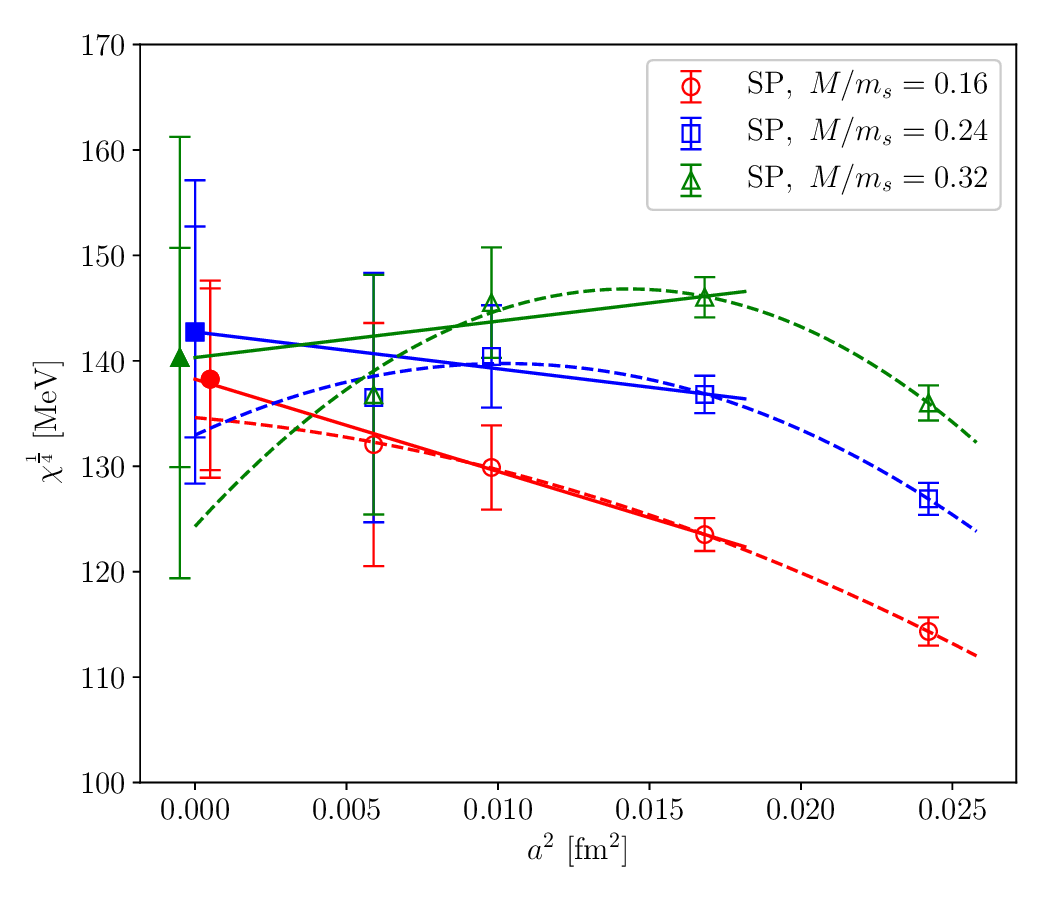}
\includegraphics[scale=0.42]{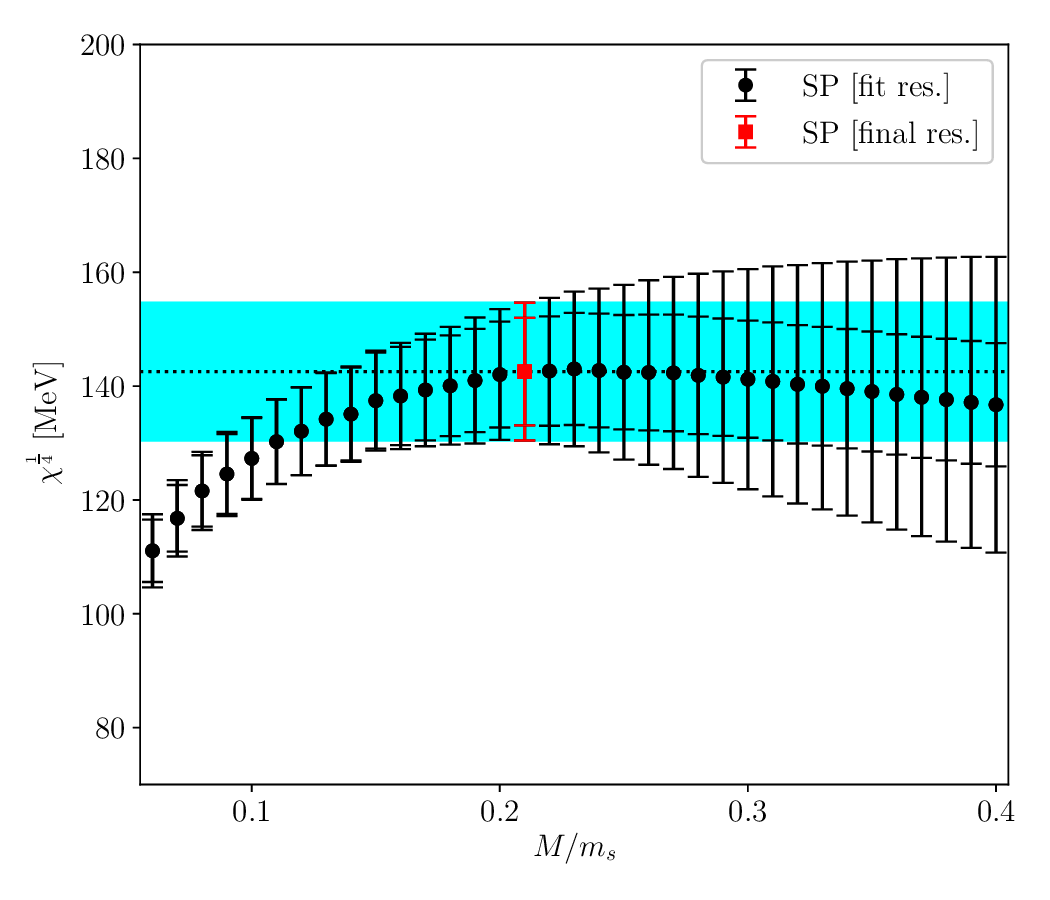}
\caption{Left panel: comparison of continuum limit extrapolations of $\chi_\SP^{1/4}(M/m_s)$ for $m_l=9~m_l^{(\mathrm{phys})}$ and for a few values of $M/m_s$. Straight lines refer to the continuum extrapolations obtained by fitting determinations obtained for the 3 finest lattice spacings with a linear function in $a^2$. Dashed lines refer to the best fit result with a $a^2$-parabolic fit function. Full points in $a=0$ represent the corresponding continuum extrapolations. Right panel: continuum limits of $\chi_{\SP}^{1/4}$ obtained from spectral
projectors at $m_l=9~m_l^{(\mathrm{phys})}$ for several values of $M/m_s$. The square point and the shaded area represent our final result for the continuum limit of $\chi_\SP$. The double error bar convention is the same of the previous plots.}
\label{fig:comparison_spectral_topsusc_9ml}
\end{figure}

In this appendix, we collect additional plots not shown in the main text. In Fig.~\ref{fig:fit_mode_number_appendix}, we show the linear fit to the mode number for all lattice spacings but the finest (which is reported in the main text), and for all LCPs. In Figs.~\ref{fig:comparison_spectral_topsusc_6ml} and~\ref{fig:comparison_spectral_topsusc_9ml} we show the extrapolation towards the continuum limit of the topological susceptibility for the LCPs corresponding to $m_l = 6,9$ $m_l^{\phys}$.

\FloatBarrier

\FloatBarrier

\providecommand{\href}[2]{#2}\begingroup\raggedright\endgroup

\end{document}